# Comparison of OpenAlex and Scopus coverage of German institutions' publications in top-tier journals

Andrey Lovakov*, Ivan Sterligov**

*lovakov@dzhw.eu
0000-0001-8644-9236
German Centre for Higher Education Research and Science Studies (DZHW), Germany

**sterligov@e-kvadrat.com
0000-0001-9736-4713
E-Quadrat Science and Education, Germany

## Abstract
OpenAlex has recently emerged as a leading alternative to proprietary bibliometric sources. However, concerns remain regarding the quality of its metadata, especially the institutional profiles which are crucial for evaluating organizations. This study assesses the quality of affiliation data in OpenAlex using German research institutions. Publications from top-tier journals were analyzed and institutional publication counts in OpenAlex were systematically compared with counts in Scopus. The results show that OpenAlex generally contains more publications at the journal level, reflecting its broader coverage. However, institutional publication counts in OpenAlex are consistently lower; it contained 6.8% fewer institution-paper links in the analyzed sample. Aggregate institutional counts nevertheless showed very high rank correlation and concordance with Scopus. Discrepancies varied substantially across fields, publishers and journals. Detailed investigation identified missing affiliation information, differences in institutional boundaries, and divergent year and journal assignments, while errors present in both databases. These findings suggest that OpenAlex is suitable for comparative institutional analyses in academic research but requires further improvement in affiliation metadata before it can be used for evaluation contexts that rely on absolute publication counts.

## 1. Introduction
OpenAlex has quickly become one of the most significant open bibliometric databases available for research evaluation and scientometric analysis. Launched in January 2022 as a successor to Microsoft Academic Graph (MAG), OpenAlex has positioned itself as a leading resource for large-scale analyses of scholarly communication (Priem et al., 2022). Its open data model, broad coverage, and commitment to transparency have made it an attractive alternative to commercial subscription-based databases, which are still widely used. Recent studies have shown that OpenAlex surpasses many bibliometric databases in terms of document and journal coverage, enabling more inclusive and representative bibliometric analyses (Alonso-Álvarez & van Eck, 2025; Alperin et al., 2024; Culbert et al., 2025; Maddi et al., 2025; Nazarovets et al., 2026; Simard et al., 2024). Within three years of its launch, several major research organizations began integrating OpenAlex into their workflows. Sorbonne University officially switched from Web of Science to OpenAlex as its main bibliometric data source in 2023 (Sorbonne University, 2023). Centre for Science and Technology Studies (CWTS) released the Leiden Ranking Open Edition, based entirely on OpenAlex data (Waltman et al., 2024). In France, both the Centre National de la Recherche Scientifique (CNRS) and the French Ministry of Higher Education and Research announced strategic moves toward open bibliographic infrastructures, explicitly identifying OpenAlex as a key alternative to commercial databases (Bordignon, 2024). Overton, a platform that tracks research papers cited in policy documents, switched to OpenAlex as a main source of author affiliation and document open-access information (Overton, 2025).

However, as a relatively new database, OpenAlex should be used with caution until the quality of its data has been thoroughly assessed. Since launch in 2022, researchers have examined both the coverage and quality of OpenAlex data (Forchino & Torres-Salinas, 2026). Researchers are focusing on its geographic (Alperin et al., 2024; Maddi et al., 2025; Zheng et al., 2025) and linguistic coverage (Céspedes et al., 2025), metadata completeness and accuracy (Alonso-Álvarez & van Eck, 2025; Culbert et al., 2025; Delgado-Quirós & Ortega, 2024; Zhang et al., 2024), citation indicators (Scheidsteger et al., 2025; Thelwall & Jiang, 2025), reference coverage (Cicero & Sarlo, 2026; Culbert et al., 2025), document type classification (Haupka et al., 2025), open access coverage and classification (Jahn et al., 2023; Simard et al., 2024), coverage of the retracted papers (Hauschke & Nazarovets, 2025; Ortega & Delgado-Quirós, 2024), coverage of data journals (Jiao et al., 2023). Collectively, this growing body of work highlights that while OpenAlex performs well across many dimensions, the quality of specific metadata fields varies substantially. In particular, several studies have noted that institutional affiliation data is less complete and less accurate (Alonso-Álvarez & van Eck, 2025; Zhang et al., 2024).

Among the various metadata fields, institutional affiliation data plays a particularly important role. Affiliations link publications to organizations, which is crucial for analyses that assess research output, collaboration patterns, and institutional performance. Inaccurate or incomplete affiliations can distort the representation of institutional productivity and any metrics based on publications. Consequently, assessing the completeness and precision of institutional metadata in OpenAlex is essential, especially when the database is used for evaluation or policy purposes. In the case of OpenAlex, correct document-institution links are vital at the country level also, as the countries are assigned to documents via resolved institutions. Existing research on OpenAlex's affiliation data has largely focused on the document level, reporting the share of records without institutional assignment (Alonso-Álvarez & van Eck, 2025; Forchino & Torres-Salinas, 2026; Jahn, 2025; Zhang et al., 2024). Studies that aggregate affiliations to higher levels report a more favourable picture: counts of scholars by country correlate highly between Scopus and OpenAlex, although absolute numbers differ (Akbaritabar et al., 2024). These studies have provided valuable descriptive insights but offer limited guidance for institutional users who need to know how accurately publications are attributed to specific organizations. Affiliation accuracy is lower at the institutional level than at the country level in both OpenAlex and Scopus (Mitchell et al., 2026). Previous comparisons of proprietary databases have attributed affiliation discrepancies to missing affiliations, differences in the way institutional units are grouped together, and incorrect assignments (Donner et al., 2020; Purnell, 2022). A dataset may contain affiliation data for most documents, but if a substantial share of those affiliations are misassigned or incomplete, the utility of the database for institutional assessment remains questionable. If institutional metadata are incomplete or inaccurate, the resulting analyses can misrepresent the scientific productivity and impact the results of entire organizations or regions. Understanding the extent to which OpenAlex correctly represents institutional affiliations is therefore crucial for assessing its suitability as a data source for evaluation and policy applications. However, the institution-level precision of affiliation data in OpenAlex remains largely unexplored. Most prior evaluations have treated completeness and accuracy as aggregate properties of the database, rather than examining how specific institutions are affected.

The present study aims to address this gap by assessing the quality of institutional affiliation data in OpenAlex using German institutions as a case study. To carry out this assessment, we rely on journals listed in the Norwegian Register for Scientific Journals, Series, and Publishers (the Norwegian list, https://kanalregister.hkdir.no/en). This list plays a central role in the evaluation of research in Norway (Sivertsen, 2017) and has inspired similar lists in several European countries (Aagaard et al., 2015; Pölönen, 2018). It includes journals considered to be of high academic quality, offering a balanced representation across disciplines and avoiding overreliance on citation-based metrics such as impact factors. Instead, a rigorous time-tested expert evaluation routine is used, which distinguishes it from other, metrics-based lists like CWTS Core or Chinese Academy of Sciences, and allows us to posit it as a list of sources that are deemed most reputable and important by domain experts. The list is divided into several levels, with the paper published in the top-tier awarding a university three times more points than a paper in the second tier venue.

By using the top tier of the list, we focus on works that are particularly relevant for university evaluation and reputation, where accurate institutional attribution is most consequential. The choice of this dataset serves two purposes. First, it ensures that the analyzed publications are among the most visible and influential outputs of German research institutions. Second, it serves as a proxy of how missing or incorrect affiliations could affect evaluation outcomes, since these journals tend to carry more weight in performance-based assessments. By systematically comparing institutional affiliation data in OpenAlex with corresponding data in Scopus, this paper seeks to provide a clearer picture of OpenAlex’s current strengths and weaknesses for analysis of institutions. Scopus (Baas et al., 2020) is chosen as a database that is widely used at organization level including in influential THE and QS university rankings, and in national evaluations across the World, most directly in Malaysia and Indonesia. Notably Türkiye has recently announced a switch to Scopus, implicitly mentioning its usage at the organizational level by these rankings,[1] which underscores the importance of the links we intend to explore. The findings will help clarify whether OpenAlex can already serve as a robust foundation for institutional-level research evaluations, or whether further improvement of its metadata is needed before it can replace or complement commercial databases.

# 2. Data and methods

## 2.1. Sample of journals

We start with 2032 journals from the top tier of the 2023 edition of Norwegian list. This list covers all research fields and includes 597 journals in Natural Sciences and Engineering, 533 in Humanities, 459 in Health Sciences, 443 in Social Sciences. For each journal the total number of publications in 2020-2025 were extracted from Scopus and OpenAlex. As the publication types in Scopus and OpenAlex differ a lot (Haupka et al., 2025), we count all types without filtering. The data was extracted via the Scopus and OpenAlex APIs in October 2026. To be able to isolate the issues of the data quality related with indexing the journals and the issues of the data quality related with affiliation data for analysis at the level of institutions we selected only those journals which have the same or very similar number of publications (ratio of Scopus/OpenAlex in the range 0.95-1.05) in both datasets. It means that differences in the number of publications affiliated with each organization could be interpreted as the differences in affiliation data. All journals were matched to both databases, and 639 of them were selected for the further analysis based on this criteria. Of the 639 selected journals, 630 contain at least one publication of the 200 selected German institutions.

## 2.2. Sample of institutions

We selected 200 German institutions with the highest number of publications in Scopus. For these purposes we calculated the number of articles and reviews published in journals between 2020 and 2025 for each German institution using the in-house version of the Scopus database provided by the German Competence Network for Bibliometrics (Schmidt et al., 2025).

## 2.3 Institution profiles in Scopus and OpenAlex

The processes of institution data curation and assigning paper to institution links are central to our study, and they differ per database. We focus on the uniform view of built-in institutional profiles that is most natural to the common users of the systems compared to the professional metascience practitioners working with snapshots and advanced queries.

---

[1] "Enhancing the national and international competitiveness of our universities is one of the most important missions of the Council of Higher Education. In particular, it is gratifying to see our research universities rise in global rankings. As of 2026, Scopus will be used in monitoring and evaluation processes." (Erol Özvar, President of Council of Higher Education of Türkiye) https://www.elsevier.com/about/press-releases/turkish-higher-education-council-selects-elseviers-scopus-for-national

Both databases rely on internal institution IDs and allow for complex nested organisation profiles, but, when queried directly by a parent internal ID, do not include the papers of its children if they have separate internal IDs. But these direct counts per ID are not what a user sees when they search for institutions' in OpenAlex or Scopus. Both databases instead display *lineage counts*, combining the counts of papers of a parent ID with those of all its children, but they differ in their inner workings.

OpenAlex directly relies on ROR and its hierarchy of parent/child/related organisations and offers a convenient way to retrieve the full paper set of the parent and its children (but not 'related') via its authorships.institutions.lineage data property, which is used by the OpenAlex web UI itself (see Figure 1). So, for each organisation we manually resolve the correct main ROR and internal OpenAlex Institution ID, then construct an API query using the aforementioned property and Institution ID.

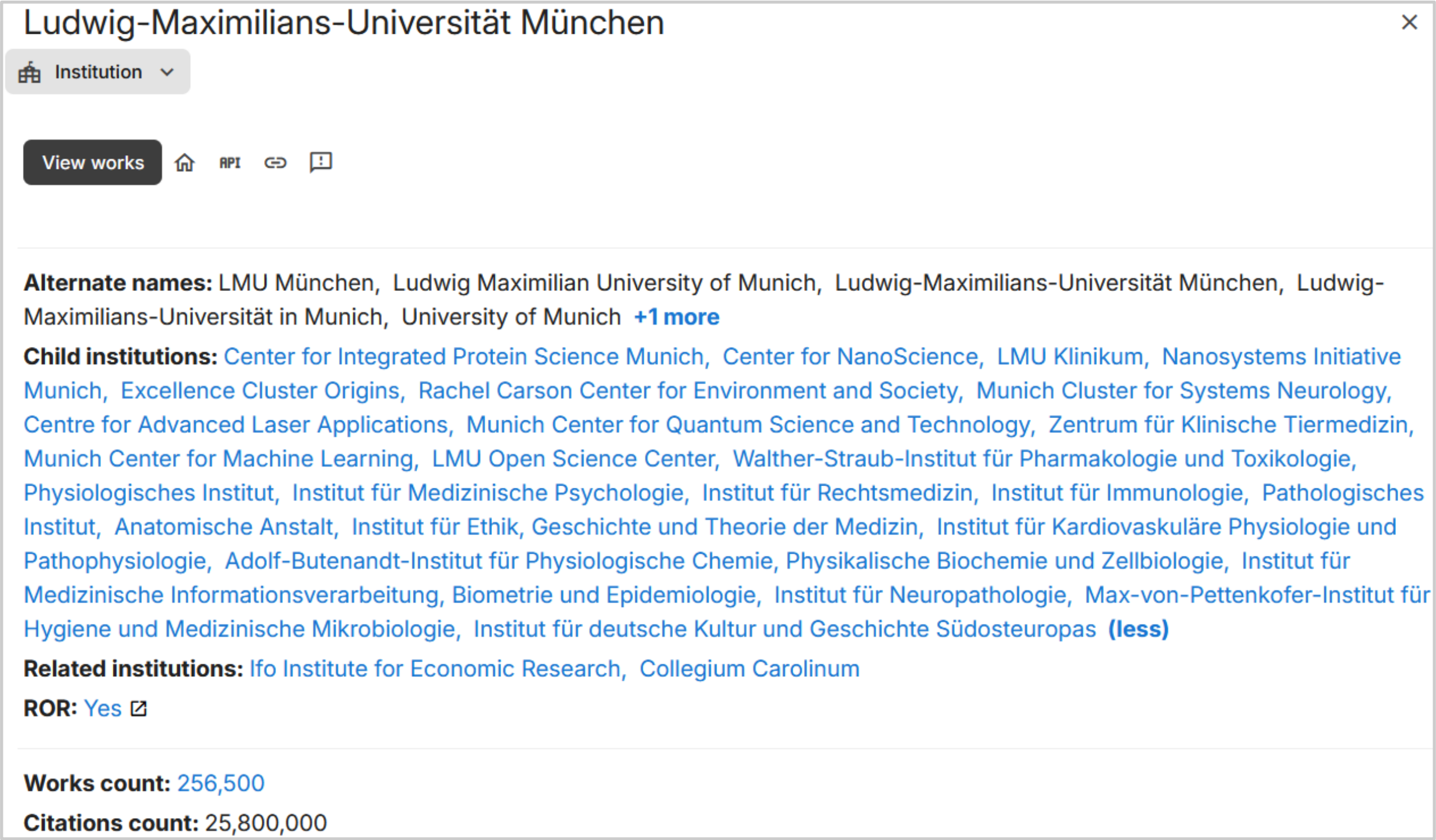

Figure 1. An example of an OpenAlex profile for a large university as seen in web UI. Clickable 'Works count' URL is
https://openalex.org/works?filter=authorships.institutions.lineage:i8204097

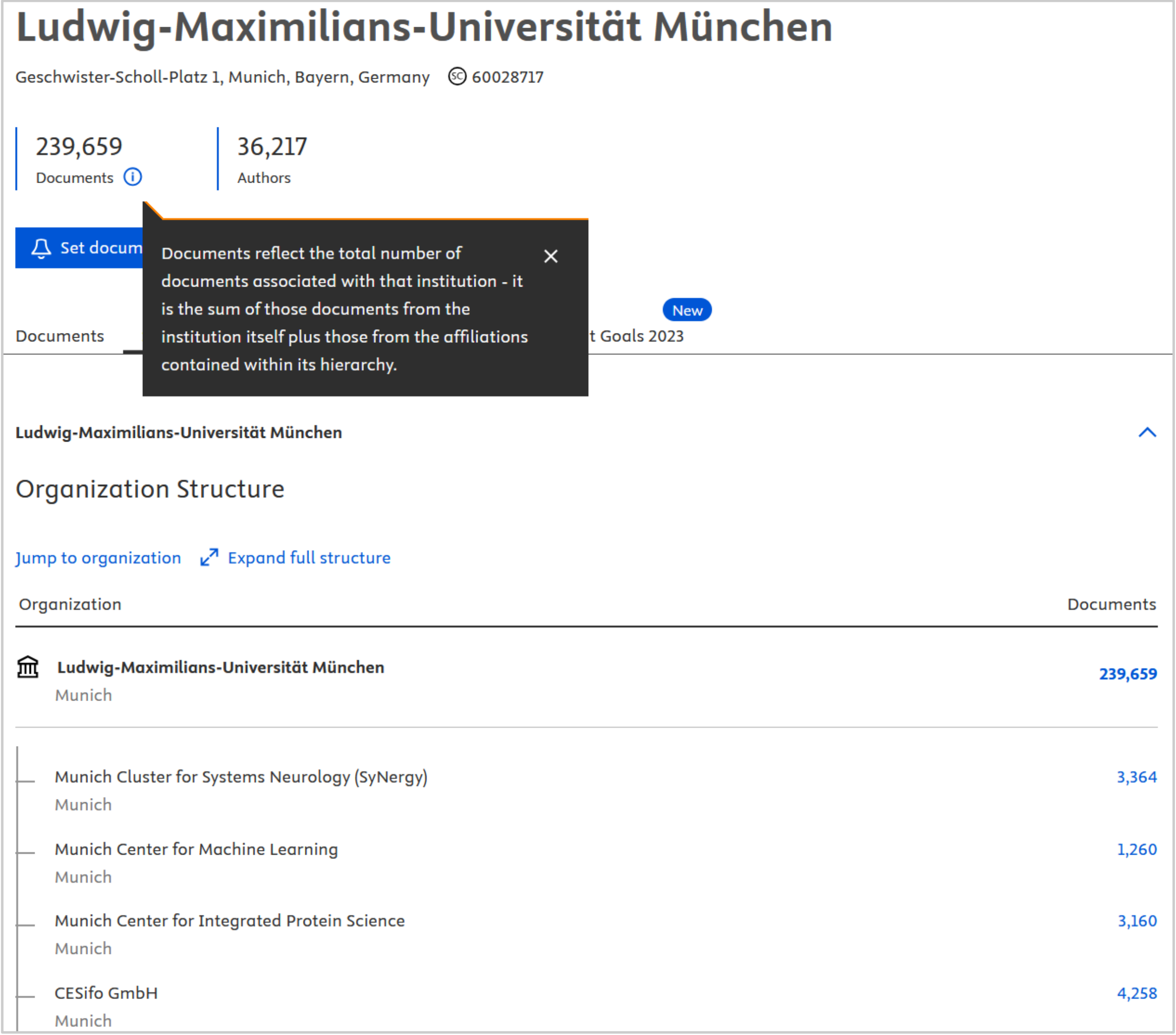

Figure 2. An example of a Scopus profile for a large university as seen in web UI. Clicking on the paper count redirects to a page with the full query for lineage.

Scopus is not based on ROR and has no mapping of its internal Affiliation IDs to it. Instead, it allows users that represent paper-producing organizations to manually add and edit children of any scale, be it a department or a single lab. Some organisations do that, while the others don't. Scopus does not offer a way to programmatically get the papers of a parent and its children by the parent's ID, so instead we need to first get the IDs of all the children, which is also not straightforward and not possible via API. First, a user has to locate and visit the main organisation profile page, then click on the total number of documents, and then the advanced search for this paper set is displayed, listing all the affiliation ids (see Figure 2). So, for an organization with the main ID 60028717 (Ludwig-Maximilians-Universität München) we get this lineage query:
AF-ID(60028717) OR AF-ID(60117647) OR AF-ID(60118149) OR AF-ID(60078288) OR AF-ID(60277819) OR AF-ID(60136274) OR AF-ID(60117698) OR AF-ID(60003930) OR AF-

ID(60122653) OR AF-ID(60006010) OR AF-ID(60000291) OR AF-ID(60098833) OR AF-ID(60075024) OR AF-ID(60280918) OR AF-ID(60277753) OR AF-ID(60277750) OR AF-ID(60170301) OR AF-ID(60103691) OR AF-ID(60103688) OR AF-ID(60072345). It yields 239k documents (totals, as of February 2026), while the query AF-ID(60028717) yields 202k. We manually construct such queries for all the organizations, of which 92 lack children IDs and 14 have 10+ children each. As the two databases use different sets of parent/child hierarchies, we expect this to be one of the causes of discrepancies in lineage paper counts for larger universities, i.e. for those with affiliated hospitals (Elizondo et al., 2022).

### 2.4. Data collection

For each of 200 selected German institutions, information about the number of affiliated publications published between 2020 and 2025 in selected journals were extracted from Scopus and OpenAlex databases. The data was extracted in October 2026 via official Scopus and OpenAlex APIs using aforementioned lineage queries that were combined with internal journal IDs for selected journals and years. The data, the Online Supplement, and the code for data collection and analysis are available at https://github.com/IvanSterligov/Oalex_vs_Scp_orgs_journals and can be rerun, as both Scopus and OpenAlex offer free API tiers, it is optimized for maximum speed. To ensure better data quality, instead of relying on ISSNs we used internal source IDs for Scopus and OpenAlex that were matched with the journals. We confirm that the important earlier limitation of OpenAlex (inability to return papers by institution ID if an affiliated author is not in the first 100 authors, mentioned in the earlier conference paper) is rectified in 2026. We use the API-based approach instead of relying on data dumps because we aim to make a reproducible contribution or, rather, a rerunnable comparison, and to use the same data collection scenarios and tooling that are available and widely used by various bibliometrics/evaluation practitioners and regular OpenAlex users. Thus instead of datasets that require access and complex setup, we put effort into making a fast and free way to reproduce (and update to monitor changes) our dataset for everyone. Still, it is important to note that our API-based approach makes byte-for-byte replication impossible due to constant data changes.

### 2.5. Data analysis

To compare the publication counts of research organizations across two databases we aggregated the data at five distinct levels of granularity:

- Organization level. The full dataset contained publication counts per organization, year, journal, and database. We first grouped the records by organization identifiers. For each organization, we computed the total number of publications indexed in OpenAlex and in Scopus across all years and journals.
- Organization-year level. To investigate temporal variation, we grouped the data by organization and publication year. For each organization-year combination, we summed the publications in OpenAlex and Scopus separately.

- Organization-field level. To assess the differences between research fields, we aggregated the number of publications by organization and research fields (each journal is assigned to one research field in the Norwegian list). The number of publications in OpenAlex and Scopus were calculated separately for each organization-field pair.
- Organization-publisher level. To assess the influence of the publisher, we grouped the data by organization and publisher (as identified in Scopus metadata). Publication counts in OpenAlex and Scopus were aggregated separately for each organization-publisher pair.
- Journal level. To identify the journals in which the differences are higher, we aggregated the publication of all selected institutions in OpenAlex and Scopus separately for each journal.

For each aggregation level, two comparative measures were calculated:

$$\text{Raw difference: } diff = OpenAlex_{ij} - Scopus_{ij}$$

$$\text{Relative difference: } reldiff = \frac{OpenAlex_{ij} - Scopus_{ij}}{OpenAlex_{ij} + Scopus_{ij}}$$

where $OpenAlex_{ij}$ is the number of publications for organization or journal $i$ in the aggregation unit $j$ (year, field, publisher, or overall) in OpenAlex, and $Scopus_{ij}$ is the corresponding number of publications in Scopus.

The raw difference represents the absolute difference in the number of publications between the two databases, while the relative difference normalizes this difference by the total number of publications indexed in both databases for the same aggregation unit, providing a scale-independent comparison. The relative differences range between −1 (publications only in Scopus) and +1 (publications only in OpenAlex).

Following Scheidsteger et al., (2025), we calculated two types of correlation coefficients to assess the similarity between two databases: 1) Spearman's rank correlation coefficient, and 2) Lin's concordance correlation coefficient, which combines measures of precision and accuracy to determine how far the observed data points deviate from the line of perfect concordance (the 45-degree line on a square scatter plot).

For the comparison of the subgroups (years, fields, and publishers) three indicators were also used:

- The percentage difference on summed counts: $PercPooledDiff = 100 * \frac{\sum OpenAlex_{ij} - \sum Scopus_{ij}}{\sum Scopus_{ij}}$, which shows how many percent more (positive) or fewer (negative) institution-paper links OpenAlex contains than Scopus.
- The shares of institutions whose OpenAlex number of publications is more than 10% above their Scopus number, within ±10% of it, or more than 10% below it.

- The 10th and 90th percentiles of the institutions' OpenAlex/Scopus ratios, expressed as percentage differences: 100 * (ratio − 1).

These three indicators were calculated over institutions with at least 10 publications in the subgroup, because for institutions with only a few publications a single publication changes the ratio dramatically. Fields and publishers are shown separately only if they include publications of at least 30 institutions in at least 3 journals.

# 3. Results

## 3.1. Comparison of journal-level data

We first compared the total number of publications across journals in Scopus and OpenAlex. Figure 3 presents the relative differences for 2,032 journals. OpenAlex contained more publications than Scopus for 81% of the journals, Scopus for 19% and 1% of the journals had equal numbers. The median relative difference was 0.045 (IQR [0.007, 0.141]), and median raw difference was 54 publications (mean = 477.4). The distribution of relative differences was skewed (Figure 3), with several journals showing particularly large discrepancies. The journal number of publications in the two databases were strongly correlated in rank (Spearman's $\rho = 0.948$), but their absolute agreement was low (Lin's CCC = 0.770).

Tables 1–2 list the 20 journals with the largest absolute differences in publication counts between the two databases for the years 2020–2025. These cases illustrate that the highest discrepancies are within the medical/health sciences as almost all top journals with the highest differences are from this area. A closer inspection suggests several underlying causes. First, OpenAlex indexes supplementary issues with conferences' papers, abstracts and posters for some journals that are not covered in Scopus, leading to higher publication counts. Second, differences in the assignment of publication year contribute to mismatches within the analyzed time window. OpenAlex generally records the earliest available publication year (e.g., online first), whereas Scopus assigns the year corresponding to the formal issue publication. As a result, some articles fall within the 2020–2025 range in one database but outside it in the other. Third, certain document types, such as book reviews, biographies, obituaries, tributes, and interviews are indexed in OpenAlex but are not consistently included in Scopus, further increasing the divergence in counts. Finally, discrepancies are also caused by isolated metadata errors in OpenAlex, such as incorrect publication years or sources.

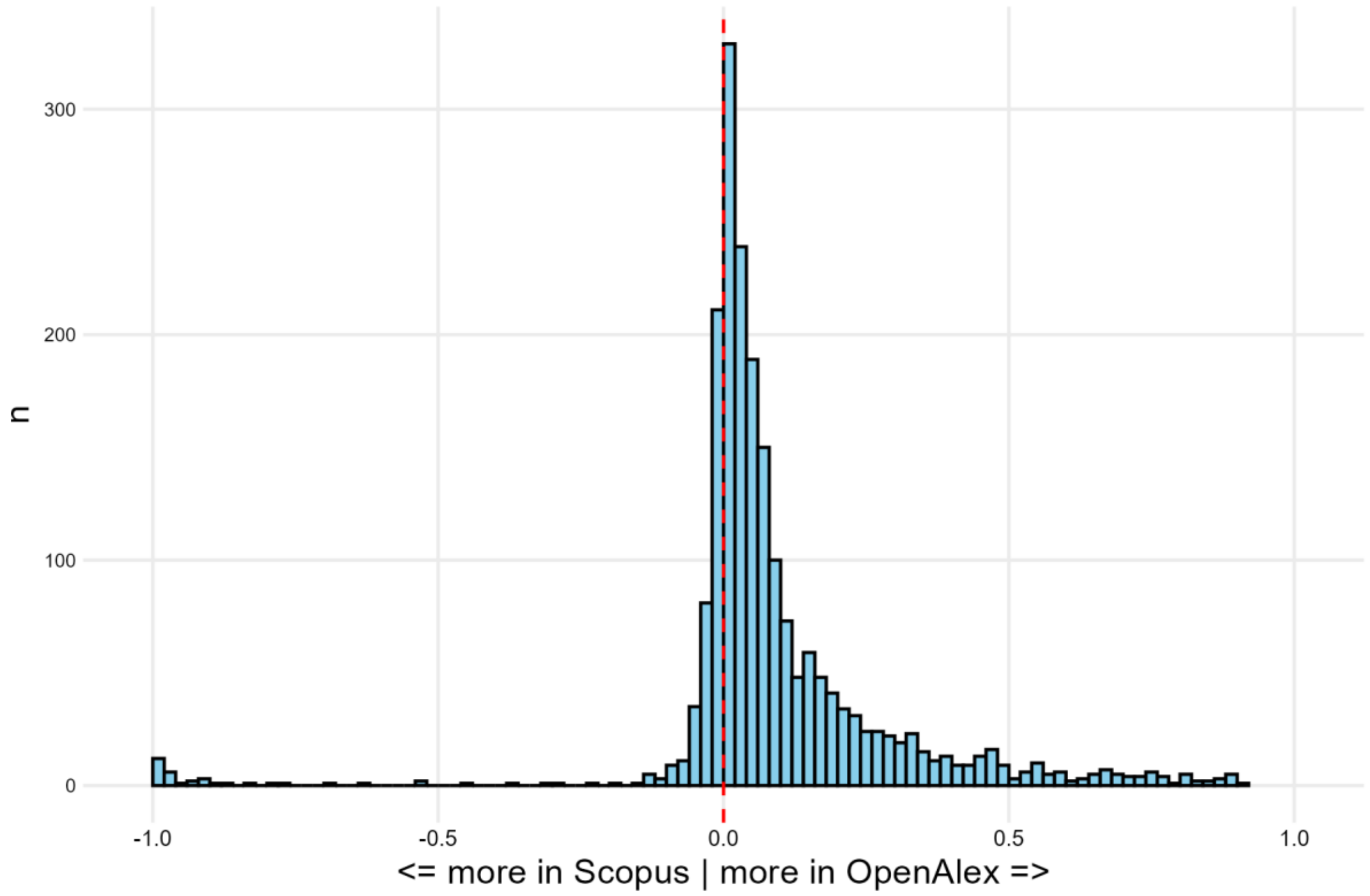

Figure 3. Relative difference in total number of publications 2020–2025 for 2032 journals in Scopus and OpenAlex.

Table 1. Journals with the largest differences in number of publications between Scopus and OpenAlex, 2020–2025, more in OpenAlex

| **Title** | **OpenAlex** | **Scopus** | **diff** |
|---|---|---|---|
| Cancer Research | 48362 | 2562 | 45800 |
| Journal of Clinical Oncology | 46814 | 7600 | 39214 |
| Blood | 44329 | 5371 | 38958 |
| Journal of the American College of Cardiology | 35380 | 4517 | 30863 |
| Circulation | 30822 | 3939 | 26883 |
| American Journal of Gastroenterology | 31147 | 6506 | 24641 |
| Annals of Oncology | 25359 | 1262 | 24097 |
| Gastroenterology | 28025 | 3942 | 24083 |
| Value in Health | 24822 | 1325 | 23497 |
| European Heart Journal | 28733 | 5573 | 23160 |
| Journal of the American Society of Nephrology | 24623 | 1740 | 22883 |
| Chest | 23750 | 4231 | 19519 |
| Annals of the Rheumatic Diseases | 21637 | 2323 | 19314 |
| Neurology | 23759 | 7306 | 16453 |
| The FASEB Journal | 20527 | 5254 | 15273 |
| Journal of Urology | 20444 | 5321 | 15123 |
| eLIFE | 22563 | 7745 | 14818 |
| Neuro-Oncology | 14768 | 1617 | 13151 |
| British Journal of Surgery | 15552 | 2776 | 12776 |
| Medicine & Science in Sports & Exercise | 15628 | 2979 | 12649 |

Table 2. Journals with the largest differences in number of publications between Scopus and OpenAlex, 2020–2025, more in Scopus

| Title | OpenAlex | Scopus | diff |
|---|---|---|---|
| New England Journal of Medicine | 6622 | 8575 | −1953 |
| Journal of machine learning research | 91 | 2022 | −1931 |
| Science of the Total Environment | 43458 | 44840 | −1382 |
| Zeitschrift für Papyrologie und Epigraphik | 11 | 944 | −933 |
| Journal of Cleaner Production | 25973 | 26895 | −922 |
| Chemical Engineering Journal | 46768 | 47683 | −915 |
| Synthese | 2860 | 3696 | −836 |
| Nature Nanotechnology | 1489 | 2221 | −732 |
| Distributed Computing Systems | 154 | 832 | −678 |
| International mathematics research notices | 2125 | 2795 | −670 |
| International Journal of Geographical Information Science (IJGIS) | 25 | 634 | −609 |
| Catalysis Today | 2624 | 3187 | −563 |
| Construction and Building Materials | 25446 | 25853 | −407 |
| Journal of Hazardous Materials | 18428 | 18819 | −391 |
| The Burlington Magazine | 5 | 376 | −371 |
| Clinical Infectious Diseases | 6232 | 6585 | −353 |
| Bioinformatics | 5062 | 5408 | −346 |
| Harvard Law Review | 6 | 334 | −328 |
| IEEE Transactions on Knowledge and Data Engineering | 2678 | 2991 | −313 |
| Nature Chemistry | 1619 | 1912 | −293 |

### 3.2. Comparison of institution-level data

To assess the quality of institutional affiliation data, we focused on journals with similar total publication counts in both databases (Scopus/OpenAlex ratio is between 0.95 and 1.05). Within this subset, we analyzed publication counts for 200 German institutions across 630 journals (the selected journals that contain at least one publication of these institutions). Figure 4 shows the distribution of relative differences between OpenAlex and Scopus. In contrast to the journal-level results, two thirds (68%) of the institutions had a higher number of publications in Scopus, 28% had a higher number in OpenAlex, and 3% had an equal number of publications. OpenAlex contained 6.8% fewer institution-paper links than Scopus. The median relative difference was −0.017 (IQR [−0.042, 0.002]), and the median raw difference was −10 publications (mean −79.0). This indicates that institutional publication data in OpenAlex tend to underrepresent outputs compared to Scopus. Despite these differences, the relationship between institutional publication counts in the two databases was very strong (Spearman's $\rho = 0.993$), showing that institutional rankings by publication volume are nearly identical in both sources. Lin's concordance correlation coefficient is 0.991, which could be interpreted as almost perfect agreement.

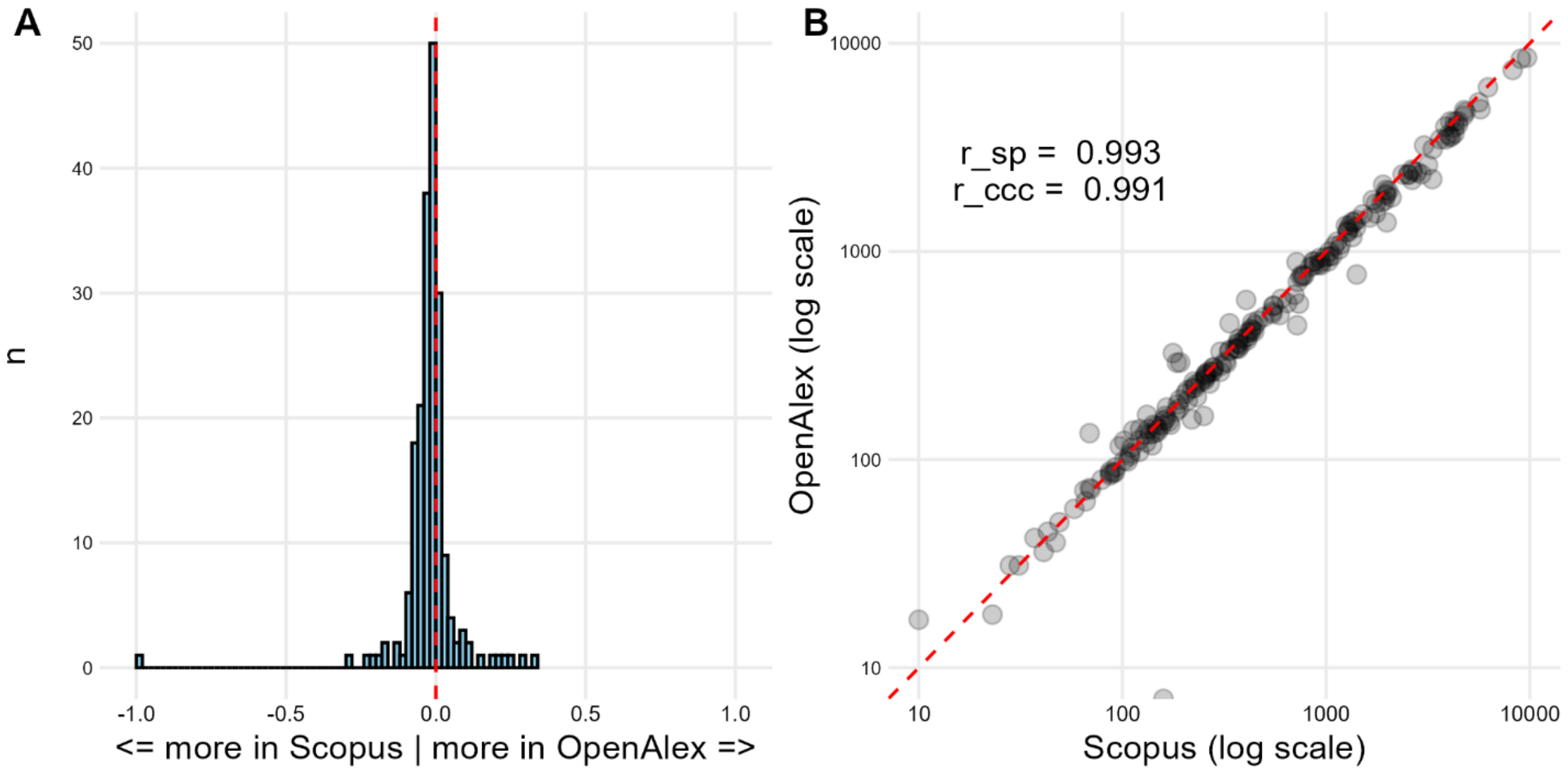

Figure 4. (A) Relative difference in total number of publications for 200 selected German institutions in 630 selected journals in Scopus and OpenAlex. (B) Total number of publications for 200 selected German institutions in both databases.

Tables 3 and 4 present the 20 institutions with the lowest and highest relative differences, respectively, between OpenAlex and Scopus for 2020–2025. Institutions from various sectors and research fields appear in both groups, suggesting that discrepancies could not be related with specific institutional types or domains.

Table 3. Top 20 institutions with the lowest relative differences in number of publications between OpenAlex and Scopus, 2020–2025

| Name | OpenAlex | Scopus | diff | reldiff |
|---|---|---|---|---|
| Max-Planck-Institut für Geoanthropologie | 0 | 159 | −159 | −1.000 |
| Universität zu Lübeck | 775 | 1414 | −639 | −0.292 |
| Brandenburgische Technische Universität Cottbus-Senftenberg (BTU) | 442 | 720 | −278 | −0.239 |
| Senckenberg Forschungsinstitut und Naturmuseum Frankfurt | 162 | 251 | −89 | −0.215 |
| Friedrich-Schiller-Universität Jena | 2226 | 3322 | −1096 | −0.198 |
| Martin-Luther-Universität Halle-Wittenberg | 1379 | 1983 | −604 | −0.180 |
| Leibniz-Institut DSMZ-Deutsche Sammlung von Mikroorganismen und Zellkulturen GmbH | 156 | 219 | −63 | −0.168 |
| Leibniz-Institut für Festkörper- und Werkstoffforschung Dresden (IFW) e.V. | 561 | 736 | −175 | −0.135 |
| FOM Hochschule für Ökonomie und Management | 18 | 23 | −5 | −0.122 |
| Christian-Albrechts-Universität zu Kiel | 2349 | 2938 | −589 | −0.111 |
| Universität Leipzig | 2599 | 3164 | −565 | −0.098 |
| Weierstraß-Institut für Angewandte Analysis und Stochastik (WIAS) | 117 | 140 | −23 | −0.089 |
| Universität Duisburg-Essen | 2208 | 2642 | −434 | −0.089 |
| Max-Planck-Institut für Biochemie (MPIB) | 495 | 592 | −97 | −0.089 |
| Rheinische Friedrich-Wilhelms-Universität Bonn | 4822 | 5732 | −910 | −0.086 |
| Roche Deutschland Holding GmbH | 40 | 47 | −7 | −0.080 |
| Medizinische Hochschule Brandenburg Theodor Fontane | 147 | 172 | −25 | −0.078 |
| Heinrich-Heine-Universität Düsseldorf | 2408 | 2799 | −391 | −0.075 |
| Leibniz-Institut für Polymerforschung Dresden (IPF) | 200 | 232 | −32 | −0.074 |
| Freie Universität Berlin | 3687 | 4271 | −584 | −0.073 |

Table 4. Top 20 institutions with the highest relative differences in number of publications between OpenAlex and Scopus, 2020–2025

| Name | OpenAlex | Scopus | diff | reldiff |
|---|---|---|---|---|
| Universität Koblenz | 134 | 69 | 65 | 0.320 |
| Forschungszentrum Borstel - Leibniz Lungenzentrum | 325 | 177 | 148 | 0.295 |
| Technische Hochschule Ostwestfalen-Lippe | 17 | 10 | 7 | 0.259 |
| Johann Heinrich von Thünen-Institut - Bundesforschungsinstitut für Ländliche Räume, Wald und Fischerei | 292 | 185 | 107 | 0.224 |
| Bernhard-Nocht-Institut für Tropenmedizin (BNITM) | 293 | 192 | 101 | 0.208 |
| Max-Planck-Institut für Struktur und Dynamik der Materie (MPSD) | 585 | 405 | 180 | 0.182 |
| Forschungsinstitut zur Zukunft der Arbeit GmbH (IZA) | 452 | 336 | 116 | 0.147 |
| Fraunhofer-Institut für Translationale Medizin und Pharmakologie ITMP | 164 | 132 | 32 | 0.108 |
| Max-Planck-Institut für Festkörperforschung (MPI-FKF) | 889 | 716 | 173 | 0.108 |
| Deutsches Rheuma-Forschungszentrum Berlin (DRFZ) | 138 | 113 | 25 | 0.100 |
| Wissenschaftszentrum Berlin für Sozialforschung gGmbH (WZB) | 123 | 102 | 21 | 0.093 |
| Max-Planck-Institut für Mathematik | 116 | 97 | 19 | 0.089 |
| Max-Planck-Institut für Plasmaphysik (IPP) | 140 | 122 | 18 | 0.069 |
| Hochschule Osnabrück | 42 | 37 | 5 | 0.063 |
| Senckenberg Museum für Naturkunde Görlitz | 31 | 28 | 3 | 0.051 |
| Helmholtz Zentrum München - Deutsches Forschungszentrum für Gesundheit und Umwelt | 2091 | 1906 | 185 | 0.046 |
| MSH Medical School Hamburg - University of Applied Sciences and Medical University | 71 | 65 | 6 | 0.044 |
| Julius Kühn-Institut (JKI) - Bundesforschungsinstitut für Kulturpflanzen | 330 | 303 | 27 | 0.043 |
| Max-Planck-Institut für Verhaltensbiologie | 178 | 165 | 13 | 0.038 |
| Deutsches Elektronen-Synchrotron (DESY) | 3230 | 3030 | 200 | 0.032 |

We also examined changes in the differences over time (Figure 5). Every year OpenAlex contains fewer institution-paper links than Scopus, and the difference grows over the analysed period: from −4.9% in 2020 and −5.5% in 2021 to −7.6% in 2022, −7.8% in 2023 and −8.2% in 2024; in 2025 it is somewhat smaller (−6.6%). The median institution follows the same trend, with a median relative difference of −0.013 in 2020, −0.030 in 2024, and −0.019 in 2025. The share of institutions with more than 10% fewer publications in OpenAlex rises from 23% in 2020–2021 to 29–31% in 2022–2025, while the share of institutions within ±10% ranges from 60–64% in 2020–2021 to 56% in 2025. The share of institutions with more than 10% more publications in OpenAlex is between 10 and 17% in all years. This suggests that discrepancies in institutional affiliation data between the two databases have grown in recent years. However, the correlation between the number of publications in the two databases has remained high throughout. Spearman's ρ ranges from 0.984 in 2020 to 0.991 in 2024. Lin's CCC ranges from 0.988 in 2022–2024 to 0.993 in 2020, which could be interpreted as substantial agreement in every year (Appendix 1).

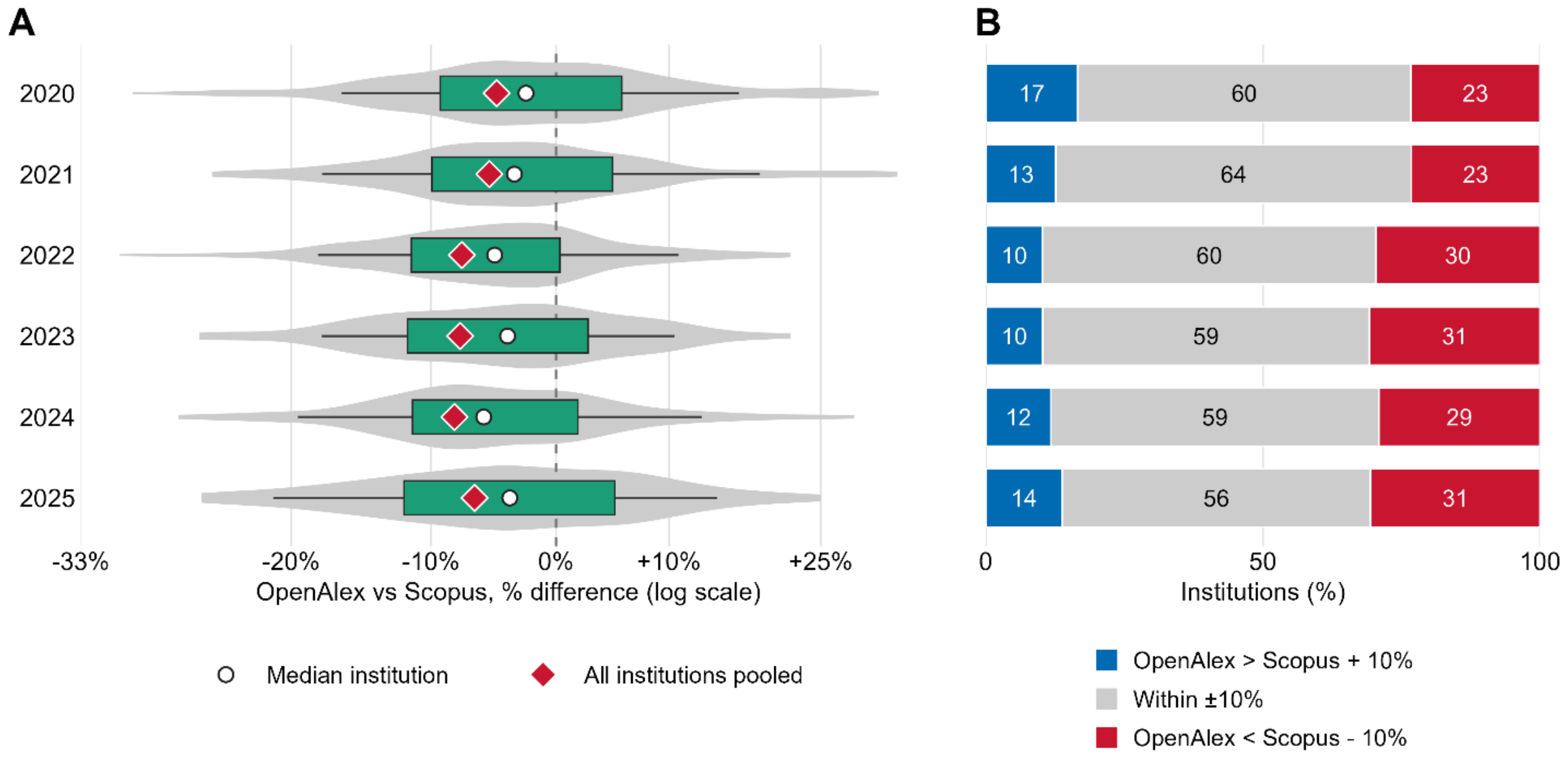

Figure 5. (A) Difference between the OpenAlex and Scopus number of publications of the institutions by year. Violin − central 90% of the institutions, green box − interquartile range, circle − median institution, whiskers − 10th–90th percentile. Diamond − the mean of the institutions' % differences weighted by their Scopus counts. (B) Share of institutions whose OpenAlex number of publications is more than 10% above, within ±10% of, or more than 10% below their Scopus number of publications.

We also analyzed discrepancies between databases across research fields. To ensure robust comparisons, we only included fields containing at least three journals and at least 30 German institutions that had published in journals categorised as belonging to those fields, according to the Norwegian list. For each field, we calculated the percentage difference on the summed number of publications and shares of institutions with more publications in OpenAlex or Scopus (see Figure 6). The results show a clear variation in the level of discrepancy across different fields of research. Institutions have more publications in Scopus than in OpenAlex in 39 of the 43 fields. The deficit in OpenAlex is smaller than 5% in 18 fields, between 5% and 10% in 12 fields, and 10% or more in 9 fields. The largest deficit is observed in Haematology (−36.3%), General Technology (−22.3%), Oncology (−20.0%), Cardiovascular and Respiratory Systems (−18.3%), Radiology, Nuclear Medicine and Medical Imaging (−17.9%), and Infectious Diseases (−17.4%) . In General Technology and most clinical fields, the majority of institutions have more than 10% fewer publications in OpenAlex (e.g. 80% Haematology, 79% in General Technology, 67% in Cardiovascular and Respiratory Systems), whereas in Earth sciences, Chemistry, Materials Science and Engineering, and Physics more than three quarters of the institutions have practically equal number of publications. Eight of the nine fields with a deficit of 10% or more are from Health Sciences. The other is General Technology (NSE). Smaller but notable deficits are found in Economics (−9.9%), Mathematics (−7.9%), Physics (−7.8%) and Business and Finance (−7.1%).This suggests that differences in data quality are not concentrated in a specific discipline, but observed across the different research fields. Differences in single years fluctuate strongly in fields with few publications and longer waiting time between publishing papers online and including them in issues, particularly in Social Sciences and Humanities. In several of these fields the differences are especially large (e.g. Political Science, Library and Information Science, Media and Communication, Education and Educational Research).

The correlation between the number of publications in the two databases is high in almost every research field. Spearman's ρ ranges from 0.996 in Chemistry and in Materials Science and Engineering to 0.513 in Media and Communication (Appendix 2). It is lower than 0.9 in three fields (Media and Communication, Sport Sciences, Haematology). Lin's CCC is below 0.90 in three fields: Haematology (0.754), Economics (0.818), Media and Communication (0.899). But overall, field level agreement remains high.

| | | 2020 | 2021 | 2022 | 2023 | 2024 | 2025 | 2020-25 | | | | Inst. | Jour. |
|---|---|---|---|---|---|---|---|---|---|---|---|---|---|
| Chemistry | Natural Sciences and Engineering | -4 | -2 | 0 | +2 | +1 | +4 | 0 | 13 | 82 | | 113 | 13 |
| Energy | Natural Sciences and Engineering | +6 | +5 | -10 | -6 | 0 | -3 | -2 | | 73 | 16 | 49 | 6 |
| Civil Engineering | Natural Sciences and Engineering | -6 | +3 | -12 | +1 | +10 | -6 | -2 | | 71 | 23 | 31 | 8 |
| Mechanical engineering | Natural Sciences and Engineering | -1 | +1 | -4 | -5 | +2 | -5 | -2 | 16 | 73 | | 45 | 11 |
| Electronics and cybernetics | Natural Sciences and Engineering | +4 | 0 | -2 | -5 | -7 | -5 | -2 | | 68 | 24 | 90 | 7 |
| Materials Sci. & Eng. | Natural Sciences and Engineering | +4 | -1 | -3 | 0 | -5 | -9 | -2 | | 79 | | 101 | 16 |
| Computer Sci. & Informatics | Natural Sciences and Engineering | +12 | +2 | -6 | -5 | -7 | -6 | -2 | | 74 | 14 | 90 | 37 |
| Interdisc. Natural Sci. | Natural Sciences and Engineering | -5 | -2 | -4 | -3 | -2 | 0 | -3 | | 76 | 14 | 168 | 6 |
| Earth sciences | Natural Sciences and Engineering | +1 | -1 | -6 | -1 | -8 | -2 | -3 | | 85 | | 100 | 27 |
| Chemical Engineering | Natural Sciences and Engineering | -5 | +6 | -8 | -3 | -1 | -10 | -4 | | 73 | 21 | 62 | 4 |
| Biotechnology | Natural Sciences and Engineering | -5 | -8 | -9 | +1 | -1 | -4 | -4 | | 71 | 24 | 38 | 4 |
| Biology | Natural Sciences and Engineering | -5 | -8 | -10 | -2 | -9 | -1 | -6 | | 74 | 15 | 136 | 47 |
| Physics | Natural Sciences and Engineering | -5 | -1 | -4 | -12 | -13 | -10 | -8 | | 77 | 17 | 109 | 21 |
| Mathematics | Natural Sciences and Engineering | -18 | -11 | -6 | -8 | 0 | -5 | -8 | | 51 | 44 | 94 | 54 |
| Electric power engineering | Natural Sciences and Engineering | +50 | -29 | -8 | +20 | -46 | -11 | -8 | | 54 | 46 | 13 | 4 |
| General Technology | Natural Sciences and Engineering | -31 | +2 | -26 | -21 | -20 | -36 | -22 | | 18 | 79 | 73 | 6 |
| Veterinary Medicine | Health Sciences | +1 | -9 | +4 | +6 | -5 | +15 | +2 | 21 | 64 | 14 | 14 | 4 |
| Endocrinology | Health Sciences | -9 | -14 | -18 | +12 | +4 | +10 | -4 | | 63 | 26 | 43 | 5 |
| Biomedicine | Health Sciences | -3 | -9 | -6 | -6 | 0 | 0 | -4 | | 71 | 21 | 130 | 39 |
| Psychology | Health Sciences | -16 | +14 | -7 | 0 | -17 | -6 | -6 | | 65 | 25 | 68 | 17 |
| Public, Environm. & Occup. Health | Health Sciences | -3 | -1 | -15 | -8 | -2 | -8 | -6 | | 64 | 27 | 90 | 22 |
| Pharmacology and Toxicology | Health Sciences | -13 | -4 | -20 | -5 | +7 | -5 | -7 | 14 | 54 | 32 | 72 | 14 |
| Sport Sciences | Health Sciences | +42 | | +7 | | -15 | -27 | -9 | 17 | 42 | 42 | 12 | 3 |
| Psychiatry | Health Sciences | +4 | -29 | -10 | -7 | +1 | -7 | -9 | 14 | 55 | 31 | 64 | 8 |
| Neurology | Health Sciences | -1 | -17 | -10 | -12 | -11 | -14 | -11 | | 54 | 41 | 76 | 7 |
| Surgical Sciences | Health Sciences | -4 | -9 | -36 | -7 | +3 | -18 | -12 | | 37 | 60 | 35 | 8 |
| General Medicine | Health Sciences | -12 | -21 | -17 | -8 | -23 | -14 | -16 | | 42 | 47 | 62 | 8 |
| Infectious Diseases | Health Sciences | -3 | -26 | +2 | -24 | -20 | -23 | -17 | | 35 | 59 | 49 | 6 |
| Radiology, Nuclear Med. & Med. Imaging | Health Sciences | -3 | -12 | -16 | -14 | -21 | -34 | -18 | | 32 | 59 | 41 | 4 |
| Cardiovascular & Respiratory Systems | Health Sciences | -11 | -10 | -28 | -24 | -7 | -25 | -18 | | 22 | 67 | 49 | 6 |
| Oncology | Health Sciences | -16 | -22 | -24 | -27 | -22 | -9 | -20 | | 29 | 61 | 49 | 7 |
| Haematology | Health Sciences | -42 | -39 | -38 | -29 | -41 | -24 | -36 | | 15 | 80 | 46 | 3 |
| Library & Information Sci. | Social Science | +38 | -22 | +53 | +16 | -24 | -4 | +7 | 24 | 68 | | 25 | 10 |
| Interdisc. Social Sci. | Social Science | -11 | +19 | -19 | +9 | +8 | -11 | 0 | 38 | 56 | | 16 | 7 |
| Education & Educational Res. | Social Science | -5 | +22 | -22 | +36 | -13 | -1 | -1 | 15 | 59 | 26 | 39 | 17 |
| Sociology | Social Science | -12 | +13 | +1 | +5 | -10 | -12 | -4 | 16 | 56 | 28 | 32 | 8 |
| Political Science | Social Science | +10 | -17 | +27 | -2 | +2 | -27 | -4 | 14 | 59 | 28 | 29 | 15 |
| Geography | Social Science | +1 | -18 | +1 | -11 | -1 | -4 | -5 | 14 | 60 | 26 | 50 | 7 |
| Business and Finance | Social Science | -2 | +7 | -5 | -17 | -7 | -18 | -7 | | 68 | 26 | 69 | 30 |
| Economics | Social Science | -9 | -4 | -9 | -14 | -13 | -11 | -10 | | 59 | 32 | 56 | 12 |
| Media and Communication | Humanities | -24 | +4 | +19 | +7 | -15 | +63 | +7 | 40 | 35 | 25 | 20 | 8 |
| Linguistics | Humanities | +20 | +3 | -14 | -2 | +20 | -26 | 0 | | 75 | 17 | 24 | 8 |
| Philosophy & History of Ideas | Humanities | +20 | +53 | 0 | -21 | -16 | -22 | -2 | 22 | 61 | 17 | 18 | 7 |

OpenAlex vs Scopus, % difference
(counts summed over institutions)

≤-30 -20 -10 0 +10 +20 ≥+30

OpenAlex > Scopus + 10%
Within ±10%
OpenAlex < Scopus - 10%

Figure 6. Left: percentage difference between OpenAlex and Scopus summed over all institutions by field and year and for 2020–2025; blue – more publications in OpenAlex, red – more in Scopus. Fields with fewer than 3 years with a value are not shown. Right: share of institutions whose OpenAlex number of publications in 2020–2025 is more than 10% above, within ±10% of, or more than 10% below their Scopus number of publications, among institutions with at least 10 publications in the field.

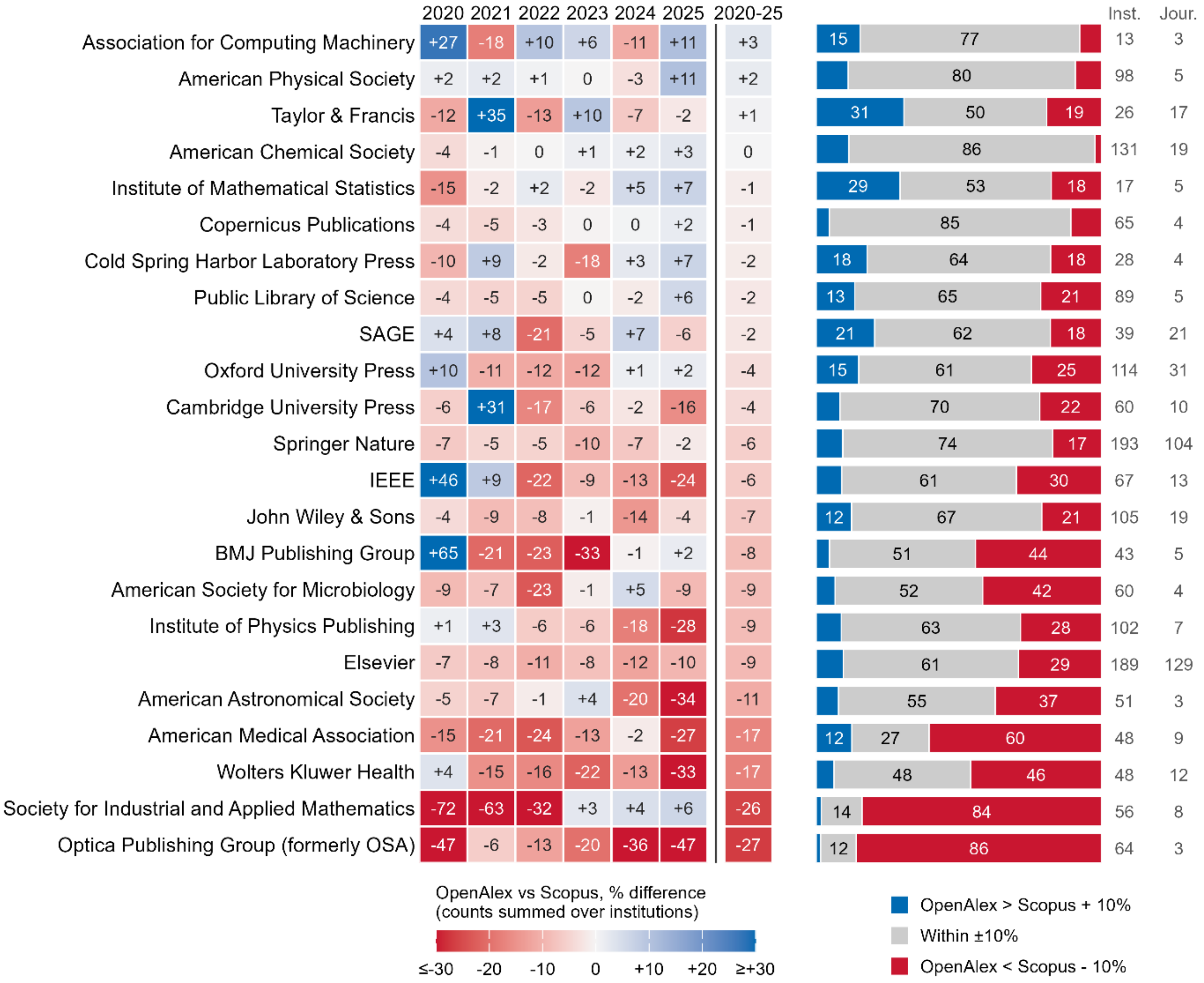

Figure 7. Left: percentage difference between OpenAlex and Scopus on the counts summed over all institutions, by publisher and year and for 2020–2025; blue – more publications in OpenAlex, red – more in Scopus, blank – fewer than 50 publications. Right: share of institutions whose OpenAlex count in 2020–2025 is more than 10% above, within ±10% of, or more than 10% below their Scopus count, among institutions with at least 10 publications by the publisher.

We also compared the differences in the number of publications between two databases in journals, published by different publishers. Institutions have a higher number of publications in Scopus for 19 of the 23 publishers (Figure 7). Only for the Association for Computing Machinery, the American Physical Society, Taylor & Francis, and the American Chemical Society OpenAlex reports slightly more institution-paper links. The largest differences are observed for Optica Publishing Group (formerly OSA) (−26.8%), the Society for Industrial and Applied Mathematics (SIAM, −26.0%), Wolters Kluwer Health (−17.5%), and the American Medical Association (−17.0%). For Optica Publishing Group and SIAM almost all institutions have more than 10% fewer publications in OpenAlex: 86% and 84%, respectively. Among the largest publishers, the difference is −9.2% for Elsevier, −6.9% for John Wiley & Sons, −5.8% for Springer Nature, whereas the numbers for the American Chemical Society (+0.2%) are almost identical. Lin's CCC is below 0.90 only for SIAM (0.861) and Optica Publishing Group (0.892). One publisher has a Spearman correlation below 0.9: Taylor & Francis (0.755). Apart from these, the databases show good agreement across publishers.

Finally, we examined the differences at the level of the journals. Of the 630 journals, 410 contain fewer institution-paper links in OpenAlex than in Scopus, 147 contain more, and 73 have equal counts. In total, 17,432 links are missing in OpenAlex and 1,632 links are found only in OpenAlex. The missing links are concentrated in a small number of journals: the 30 journals with the most missing links account for 61% of them. Physics Letters B alone accounts for 1,678 missing links (53% of its links in Scopus), followed by Astronomy and Astrophysics (1,423), Leukemia (797), Journal of High Energy Physics (657), The Astrophysical Journal (626), Optics Express (499), and European Physical Journal C (458), i.e. most journals in high-energy physics and astrophysics, where papers with very large teams from multiple organizations are common (Figure 8). Ranked by the share of missing links (Figure 8B), Physics Letters B is followed by Emerging Infectious Diseases (43%), The Lancet Haematology (41%), Reports on Progress in Physics (38%), Haematologica and Leukemia (36%), Optics Express (29%), three SIAM journals (27–29%), and JAMA Oncology (27%), which corresponds to the results at the publisher level.

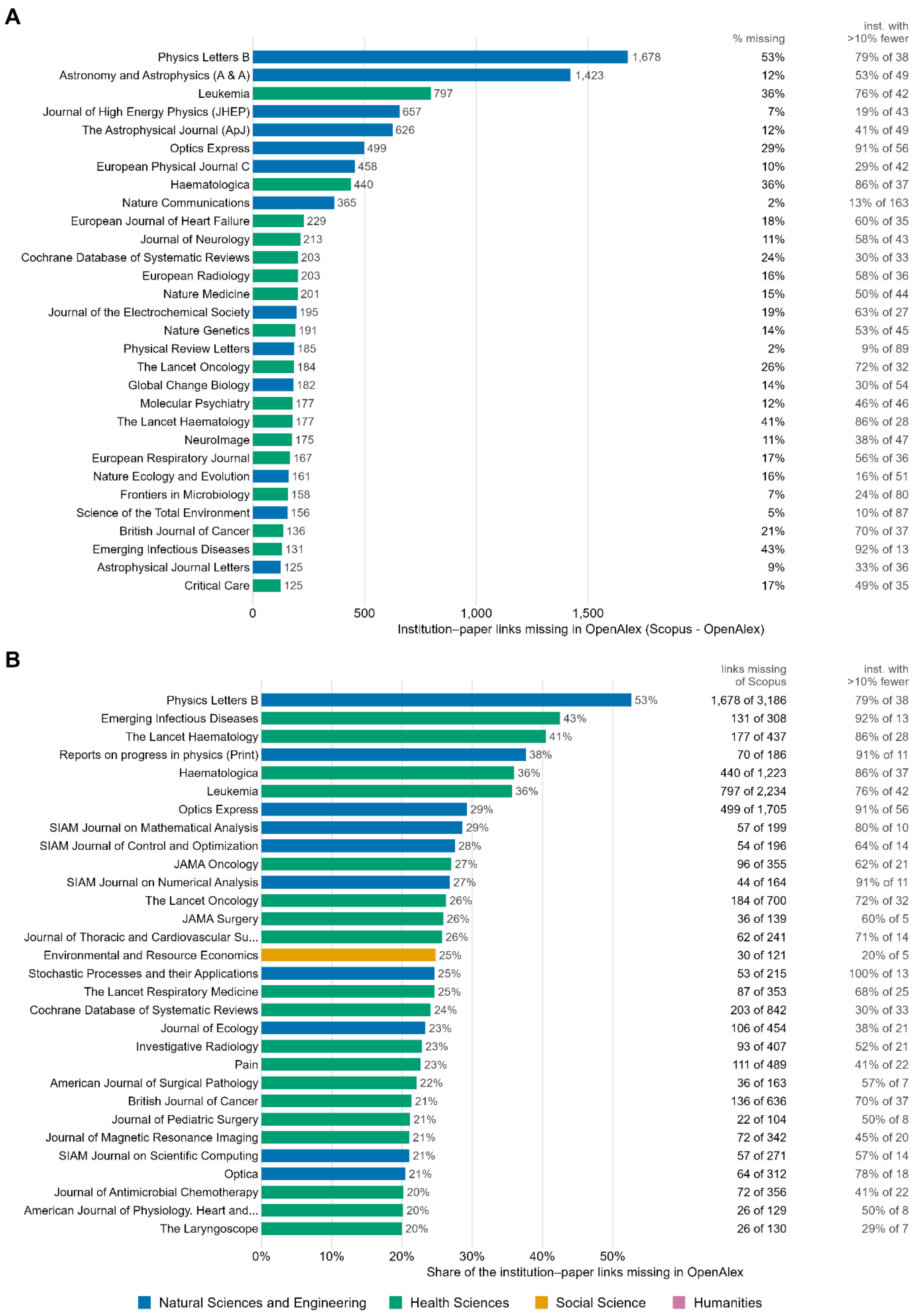
A
% missing
inst. with
>10% fewer
Physics Letters B 1,678 53% 79% of 38
Astronomy and Astrophysics (A & A) 1,423 12% 53% of 49
Leukemia 797 36% 76% of 42
Journal of High Energy Physics (JHEP) 657 7% 19% of 43
The Astrophysical Journal (ApJ) 626 12% 41% of 49
Optics Express 499 29% 91% of 56
European Physical Journal C 458 10% 29% of 42
Haematologica 440 36% 86% of 37
Nature Communications 365 2% 13% of 163
European Journal of Heart Failure 229 18% 60% of 35
Journal of Neurology 213 11% 58% of 43
Cochrane Database of Systematic Reviews 203 24% 30% of 33
European Radiology 203 16% 58% of 36
Nature Medicine 201 15% 50% of 44
Journal of the Electrochemical Society 195 19% 63% of 27
Nature Genetics 191 14% 53% of 45
Physical Review Letters 185 2% 9% of 89
The Lancet Oncology 184 26% 72% of 32
Global Change Biology 182 14% 30% of 54
Molecular Psychiatry 177 12% 46% of 46
The Lancet Haematology 177 41% 86% of 28
NeuroImage 175 11% 38% of 47
European Respiratory Journal 167 17% 56% of 36
Nature Ecology and Evolution 161 16% 16% of 51
Frontiers in Microbiology 158 7% 24% of 80
Science of the Total Environment 156 5% 10% of 87
British Journal of Cancer 136 21% 70% of 37
Emerging Infectious Diseases 131 43% 92% of 13
Astrophysical Journal Letters 125 9% 33% of 36
Critical Care 125 17% 49% of 35
0
500
1,000
1,500
Institution–paper links missing in OpenAlex (Scopus - OpenAlex)
B
links missing
of Scopus
inst. with
>10% fewer
Physics Letters B 53% 1,678 of 3,186 79% of 38
Emerging Infectious Diseases 43% 131 of 308 92% of 13
The Lancet Haematology 41% 177 of 437 86% of 28
Reports on progress in physics (Print) 38% 70 of 186 91% of 11
Haematologica 36% 440 of 1,223 86% of 37
Leukemia 36% 797 of 2,234 76% of 42
Optics Express 29% 499 of 1,705 91% of 56
SIAM Journal on Mathematical Analysis 29% 57 of 199 80% of 10
SIAM Journal of Control and Optimization 28% 54 of 196 64% of 14
JAMA Oncology 27% 96 of 355 62% of 21
SIAM Journal on Numerical Analysis 27% 44 of 164 91% of 11
The Lancet Oncology 26% 184 of 700 72% of 32
JAMA Surgery 26% 36 of 139 60% of 5
Journal of Thoracic and Cardiovascular Su... 26% 62 of 241 71% of 14
Environmental and Resource Economics 25% 30 of 121 20% of 5
Stochastic Processes and their Applications 25% 53 of 215 100% of 13
The Lancet Respiratory Medicine 25% 87 of 353 68% of 25
Cochrane Database of Systematic Reviews 24% 203 of 842 30% of 33
Journal of Ecology 23% 106 of 454 38% of 21
Investigative Radiology 23% 93 of 407 52% of 21
Pain 23% 111 of 489 41% of 22
American Journal of Surgical Pathology 22% 36 of 163 57% of 7
British Journal of Cancer 21% 136 of 636 70% of 37
Journal of Pediatric Surgery 21% 22 of 104 50% of 8
Journal of Magnetic Resonance Imaging 21% 72 of 342 45% of 20
SIAM Journal on Scientific Computing 21% 57 of 271 57% of 14
Optica 21% 64 of 312 78% of 18
Journal of Antimicrobial Chemotherapy 20% 72 of 356 41% of 22
American Journal of Physiology. Heart and... 20% 26 of 129 50% of 8
The Laryngoscope 20% 26 of 130 29% of 7
0%
10%
20%
30%
40%
50%
Share of the institution–paper links missing in OpenAlex
Natural Sciences and Engineering
Health Sciences
Social Science
Humanities

Figure 8. Journals with the most institution-paper links missing in OpenAlex, 2020–2025. (A) The 30 journals with the largest number of missing links; % missing – missing links as a share of the links in Scopus. (B) The 30 journals with the largest share of missing links; links missing of Scopus – number of missing links and number of links in Scopus. Last column in A and B (inst. with >10% fewer): share of institutions whose OpenAlex count in the journal is more than 10% below their Scopus number.

### 3.3. Possible causes of discrepancies

We don’t expect the two databases to fully agree even when all papers are correctly linked to organizations. This is because they sometimes diverge in defining what are the parts of the organization and how the entities relate. Both have “parents” and “children”, but OpenAlex relies on ROR, and Scopus relies on their own internal system (mainly curated by institution representatives themselves). In both databases children can have multiple parents, but this also differs. As OpenAlex follows external changes in ROR, it has no full control over the organizational profiles. Also, starting from 2026, paid OpenAlex customers (“members”) can directly link papers to their organizations’ profiles, while for Scopus this is since long ago available also for non-subscribers. Having admitted that, we fall short of performing the full investigation of the causes of observed differences between two databases in this manuscript at the paper level vs publisher versions of record (VoRs). We leave this complex task to the future, but instead explore at the paper level top 67 cases of major discrepancies for journal-institution pairs, including the differing profile structures.

Before we come to that, for some potential explanations it is possible to clarify the situation without comparing individual papers’ metadata. First, we examined the yearly percentage of OpenAlex works in the 639 journals studied that are not linked to any institution profile ID at all. For the vast majority, it was very low (Fig. 9), but for a group of 19 venues it is more than 50% for an average year. Almost all of them are in social sciences and humanities, and have low yearly paper counts, i.e. *Differences. A Journal of Feminist Cultural Studies* has zero institutional links since 2021. One potential explanation for the lack of affiliation data in OpenAlex is that some publishers do not share affiliation metadata via Crossref (Jahn, 2025). Where Crossref and PubMed lack affiliations, OpenAlex relies on harvesting publisher landing pages. Outside PubMed, about 40% of OpenAlex journal articles published in 2025 depend on such web-scraped affiliation data (Porter, 2026). However, this problem could potentially be solved by extracting the metadata from open-access publications, and OpenAlex actually produces and shares GROBID-extracted metadata from OA PDFs. Of the 639 journals in our sample, we selected only open-access journals according to Unpaywall. We then calculated the same indicators as for all selected journals. Summed over all institutions, OpenAlex contained 8.1% fewer institution-paper links than Scopus in open-access journals. The median relative difference was −0.017 (IQR [−0.055, 0.000]); 62% of the institutions had more publications in Scopus, 24% in OpenAlex, and 14% had equal counts.The correlations between institutional publication counts in the two databases were very strong: Spearman's ρ was 0.991 and Lin's concordance correlation coefficient was 0.990. These results suggest that the discrepancies between OpenAlex and Scopus regarding institutional publication data are not smaller but slightly larger in the sample of open-access journals.

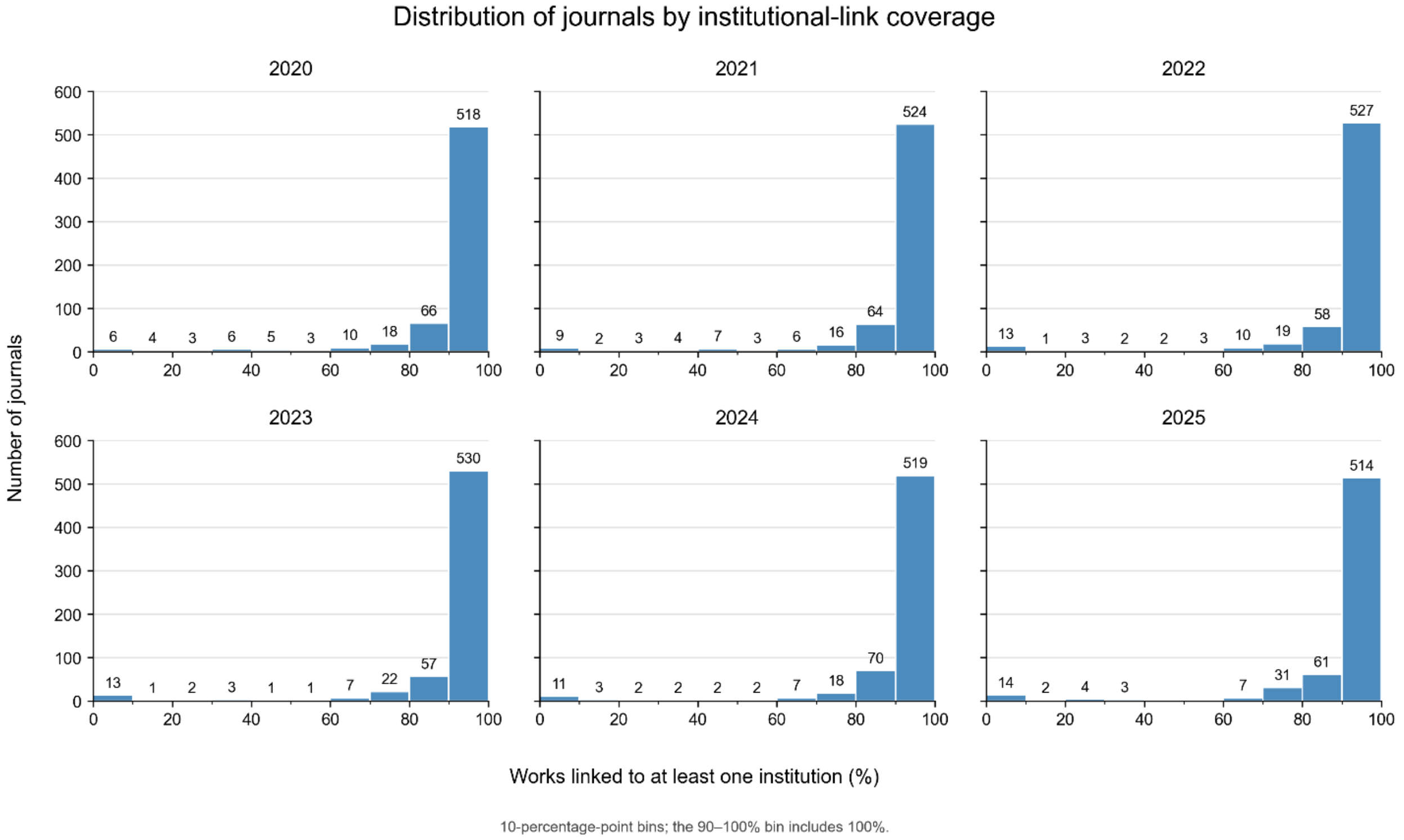

Figure 9. Yearly distribution of surveyed journals by percent of documents in OpenAlex that are linked to at least one Institution ID.

We then investigated 67 institution-journal combinations with at least 100 publications in either database in 2020–2025, and a difference of at least 30, covering 35 institutions and 14 journals (Scopus counted more in 55 combinations and OpenAlex in 12). We compared papers via DOIs and utilised paper-level and author-level affiliations, accounting for alternative names (several per institution, checked automatically, see Online Supplement), hospitals, departments and shared centres. When matching authors, we used surnames and initials. In select cases we also looked at publisher VoRs for additional evidence. We did not manually verify every paper. Examining thousands of affiliation notes is too labour-intensive and VoRs were not always accessible. The publication comparison contained 18,193 links shared by both databases, 4,645 counted for the institution only in Scopus and 1,645 only in OpenAlex. Possible discrepancies are summarised in Table 5.

Table 5. Prevalence of findings for various discrepancies in affiliation link counts for 67 institution-journal combinations in 2020–2025. Counts are institution-journal-paper links.

| *A. Paper counted for the institution in Scopus, but not in OpenAlex* | |
|---|---|
| **Finding** | **Links (count and % of total 4,645)** |
| Scopus names the institution for an author; the likely corresponding OpenAlex author entry does not retain that affiliation | 1,579 (34.0%) |
| OpenAlex provides no affiliation text for any author | 1,069 (23.0%) |
| Affiliation wording names a department or related organisation, or is ambiguous; its implications for institutional inclusion remain unresolved | 766 (16.5%) |
| Differences in the inclusion of related organisations explain the counting difference | 479 (10.3%) |
| Scopus affiliation text contains no unambiguous match to the searched institution-name variants or examined related-organisation names | 423 (9.1%) |
| Recorded year or journal differs, or the paper is in OpenAlex's XPAC expansion dataset, excluded from the searches used here | 299 (6.4%) |
| Available author information is insufficient to complete the comparison | 24 (0.5%) |
| No corresponding OpenAlex record was returned by the DOI searches | 4 (0.1%) |
| OpenAlex retains the institution's name in an author's affiliation text but does not assign the institution | 2 (0.0%) |
| ***B. Paper counted for the institution in OpenAlex, but not in Scopus*** | |
| **Finding** | **Links (count and % of total 1,645)** |
| Scopus affiliation text contains no unambiguous match to the searched institution-name variants or examined related-organisation names | 611 (37.1%) |
| Recorded year or journal assignments differ between the databases | 577 (35.1%) |
| Differences in the inclusion of related organisations explain the counting difference | 332 (20.2%) |
| Scopus provides no affiliation text for any author | 57 (3.5%) |
| No corresponding Scopus record was returned by the DOI searches | 53 (3.2%) |
| Scopus affiliation text names the institution, but exclusion from its institutional results remains unexplained | 11 (0.7%) |
| Affiliation wording names a department or related organisation, or is ambiguous, its implications for institutional inclusion remain unresolved | 4 (0.2%) |

**Institutional scope differences** arise when databases count a hospital, research centre or other related organisation differently within institutional output. The German Centre for Integrative Biodiversity Research (iDiv), for example, could contribute to a university's Scopus count while OpenAlex treated the centre separately due to it being "related" in ROR instead of "child". Conversely, OpenAlex connected Complexity and Topology in Quantum Matter Cluster (ct.qmat) to both the University of Würzburg and Technische Universität Dresden, so papers naming Dresden could enter University Würzburg's count through the cluster without a direct Würzburg affiliation.

**Paper-level affiliation missing**. OpenAlex lacked raw affiliation text for any author in 1,069 Scopus-only links. 37 Scopus papers had empty paper-level and author-linked affiliation fields, accounting for 57 OpenAlex-only links. Individual-author omissions are considered separately below.

**Year, journal and XPAC coverage**. Of 299 Scopus-only links bundled here, 258 had OpenAlex publication years outside 2020–2025, seven involved different journal assignments, and 34 concerned papers in OpenAlex’s XPAC expansion dataset, excluded from the searches used here. Of 577 OpenAlex-only links, 574 corresponded to Scopus papers dated 2026 and three involved different journal assignments. Identical year limits therefore did not ensure identical publication sets.

**Unsuccessful DOI searches**. No Scopus counterpart was returned for 53 OpenAlex-only links (43 papers), and no OpenAlex counterpart for four Scopus-only links (four papers).

**Author-specific affiliation comparisons**. We compared author entries within the same paper by surname and given-name initial, accepting only a single candidate. In 338 links, OpenAlex retained another affiliation also listed for that author in Scopus. In 642, it listed other affiliations that could not be corroborated in the Scopus entry. In 599, it had no affiliation text, although other authors on the paper did. For six selected papers, comparison with the published articles confirmed that OpenAlex had omitted an author’s institutional affiliation or recorded an incorrect one. Further article checks found errors in either database. In 10.1016/j.physletb.2020.135328 the published author list places A. Lelek at Universiteit Antwerpen, with no additional affiliations. OpenAlex instead gives the Institute for Nuclear Problems in Minsk. Scopus correctly names Antwerp but also assigns Deutsches Elektronen-Synchrotron (DESY), which the VoR does not support. By contrast, for A. Saggio in 10.1016/j.physletb.2020.135502, OpenAlex’s Université Catholique de Louvain affiliation agrees with the article, while the Scopus assignment to DESY does not. These examples show that a disagreement with Scopus cannot by itself establish an OpenAlex error. Two further links retained DESY in an author’s OpenAlex text but lacked an institutional assignment. Another 24 had insufficient author correspondence or information.

**Institution named in Scopus but not counted**. For eleven OpenAlex-only links, Scopus affiliation text named the institution, but its exclusion from Scopus’s institutional results remained unexplained.

**Related or ambiguous affiliations.** For 766 Scopus-only and four OpenAlex-only links, addresses named departments, shared centres or related organisations, or mentioned the institution in an unclear context. These suggested an institutional connection, but did not establish whether the author’s affiliation should contribute to the institution’s publication count.

The remaining 423 Scopus-only and 611 OpenAlex-only links could not be resolved despite systematic searches for institutional name variants and related organisations in the Scopus affiliation text. The searches included several German, English and shortened name variants for each institution, but standalone acronyms were excluded.

Overall, detailed comparison identified several sources of discrepancy, although some cases remained unresolved. Among Scopus-only links, 34.0% involved a likely corresponding OpenAlex author entry that did not retain the institutional affiliation recorded in Scopus, while 23.0% involved OpenAlex records with no raw affiliation text for any author. Among OpenAlex-only links, 37.1% remained unresolved because searches of Scopus affiliation text found no unambiguous match to the institution or examined related organisations. This does not establish an author-specific affiliation omission in Scopus. A further 3.5% involved Scopus records with no affiliation text for any author. Differences in institutional boundaries and parent/child/related relationships explained 10.3% of Scopus-only links and 20.2% of OpenAlex-only links. Differences in recorded year or journal accounted for 35.1% of OpenAlex-only links, while year/journal differences and XPAC exclusions together accounted for 6.4% of Scopus-only links.

## 4. Discussion

This study aimed to evaluate the quality of institutional affiliation data in OpenAlex by comparing it with Scopus, focusing on German research institutions and publications in top-tier journals. The results revealed OpenAlex's strengths and limitations, highlighting a distinction between journal-level coverage and institution-level attribution accuracy.

At the journal level, OpenAlex consistently has more publications per journal than Scopus. This finding aligns with previous research showing that OpenAlex provides broader document and journal coverage than other bibliometric databases (Alonso-Álvarez & van Eck, 2025; Alperin et al., 2024; Céspedes et al., 2025; Cicero & Sarlo, 2026; Jiao et al., 2023; Maddi et al., 2025; Zheng et al., 2025). The particularly large discrepancies observed in medical and health sciences journals further support previous evidence (Haupka et al., 2025) that OpenAlex is highly inclusive of diverse document types. However, the institution-level analysis reveals a different pattern. When focusing on journals with comparable overall publication counts in both databases, two thirds of German institutions have a higher number of publications in Scopus than in OpenAlex. This systematic underrepresentation of institutional output in OpenAlex suggests that broader journal coverage does not automatically lead to reliable institutional attribution. In other words, while OpenAlex may index more publications overall, many of these publications are either missing institutional affiliations or are not correctly linked to specific institutions. This finding directly supports earlier concerns raised in the literature regarding the lower completeness and accuracy of affiliation metadata in OpenAlex (Alonso-Álvarez & van Eck, 2025; Zhang et al., 2024).

Despite these quantitative differences, the very high correlations between institutional publication counts in OpenAlex and Scopus suggest that the institutions' relative positions remain largely the same. Based on the analysed sample of journals, institutional rankings by publication number are very close in the two databases, even though the absolute numbers differ. A similar pattern has been observed at the country level for scholarly migration (Akbaritabar et al., 2024) and for university rankings (Eck et al., 2024; Waltman et al., 2024). This suggests that, with the caveat that users are aware of its tendency to undercount publications at the institutional level, OpenAlex may already be suitable for analyses that rely on relative comparisons, such as ranking institutions by output. At the same time, the underrepresentation observed here raises concerns in evaluation contexts where absolute publication counts matter, such as performance-based funding schemes or institutional monitoring.

Temporal analysis adds an important dimension to this discussion. Before the August-September 2026 update of OpenAlex affiliation matching algorithms the discrepancies increased between 2020 and 2024, with a drop of linked papers in 2024. After the update the situation has improved markedly, with recent years having basically the same share of linked papers (Figure 10). This means not only that OpenAlex is now quite close to Scopus for our set of organizations+journals, but also that affiliation-based analyses using earlier OpenAlex snapshots that rely on institutional assignments that have since undergone substantial correction should be reassessed using the updated data. OpenAlex team highlights this in bold letters when describing the matcher update: “If you keep a copy of OpenAlex, reload it in full” (https://github.com/ourresearch/openalex-affiliation-matcher#what-changed-in-openalex).

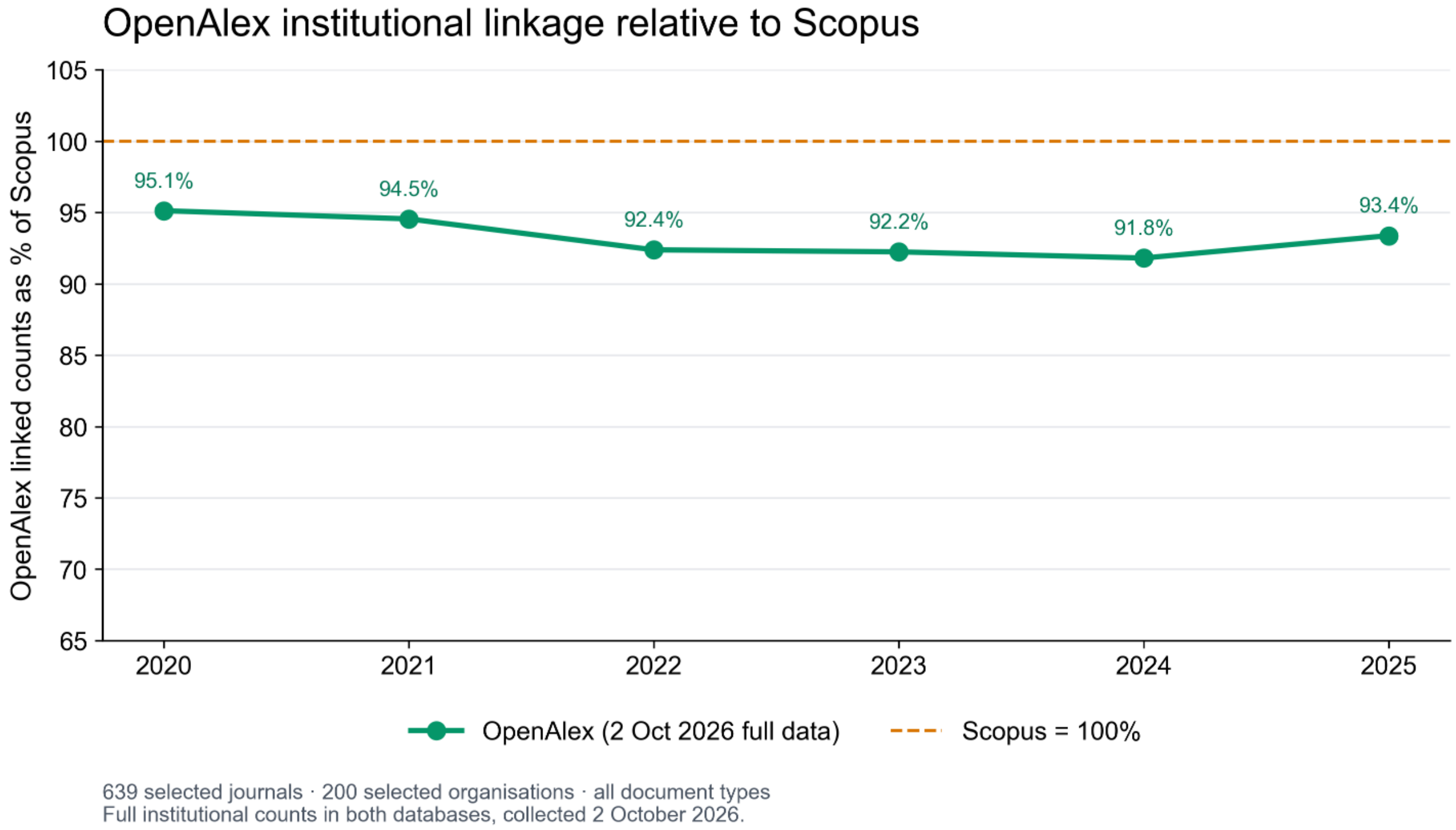

Figure 10. Yearly counts of total OpenAlex links for selected organisations+journals, as % of same counts in Scopus

Field-level analyses further demonstrate that discrepancies in institutional affiliation data are not discipline-specific. Although the differences between OpenAlex and Scopus are small in many fields, OpenAlex contains more institution-papers links than Scopus in only four of the 43 fields. The largest discrepancies are observed in fields spanning multiple disciplinary areas, including health sciences, natural sciences and engineering. The absence of a clear disciplinary pattern indicates that the observed differences are likely driven by structural issues in affiliation data processing (see Section 3.3) rather than by field-specific issues.

The publisher-level analysis reveals another potential source of variation. For most of the major publishers analysed, the number of institutional publications is higher in Scopus. For some publishers, OpenAlex shows fewer publications for almost all institutions. These differences may be related to how affiliation metadata are structured, standardized, or transmitted by publishers, and how effectively this information is parsed and linked to institutional identifiers in OpenAlex. Together with previous findings (de Jonge & Kramer, 2026; Jahn, 2025; Porter, 2026), these results suggest that publisher-specific workflows could play a role in shaping affiliation data quality and require further research.

Detailed investigation of 67 institution-journal combinations shows that differences in institutional publication counts arise through several mechanisms. Missing affiliation information is prominent among Scopus-only links. OpenAlex sometimes lacked affiliation text for the entire paper or did not have an institutional affiliation recorded for a likely same author in Scopus. These findings distinguish incomplete affiliation metadata from errors in string matching. Selected checks against published articles confirmed errors in both databases, demonstrating that disagreement with Scopus cannot itself establish an OpenAlex error. Because we targeted combinations with large discrepancies, the findings should not be interpreted as database-wide error rates.

Differences in institutional definitions also contributed to discrepancies in both directions. Scopus uses its own institutional profiles, whereas OpenAlex relies on ROR relationships to construct institutional lineages. Hospitals, shared research centres and other associated organisations can therefore contribute differently even when their affiliations are correctly identified. These differences concern the operational definition of an institution as well as metadata quality. Explicitly defining which organisational units should contribute to an institution's output is thus essential for comparisons across databases. Differences in publication-year and journal assignments provide a further explanation, illustrating why similar journal totals and identical year limits do not ensure identical publication sets. Previously documented limitation affecting searches for affiliations beyond the first 100 authors has been corrected in OpenAlex. Institutional and lineage filters now search all retained authorships, although list responses still display only the first 100.

Taken together, our results highlight the distinction between completeness and precision. These are two separate dimensions of data quality, both of which matter for institutional-level bibliometric usage. OpenAlex performs well in terms of coverage, but its current affiliation data appear less precise than those in Scopus when evaluated at the level of individual institutions. This has direct implications for research evaluation and policy use. Analyses that depend on relative comparisons across institutions may yield similar conclusions regardless of the database used. However, analyses relying on absolute publication numbers or fine-grained attribution may be more sensitive to the incompleteness of affiliation data in OpenAlex. For applied/administrative workflows that require complete coverage, i.e. CRIS and external national evaluations, the observed gaps are very important and OpenAlex lags behind Scopus in this regard.

Overall, this study suggests that OpenAlex is already a strong and promising foundation for institutional-level bibliometric analysis, but so far only for a subset of leading journals, and with large discrepancies for some of them. We suspect ours to be a more positive picture than on average for all sources and organisations, but even for the central sources such an analysis requires selecting sources with better metadata quality, and exploring the diverging children/relationships data so that it is aligned with the analysis aims and scope. It is vital to supplement our analysis with the one focused on less established sources and organisations from developing countries, which could yield very different results. We conclude with the database-wide Figure 11 that highlights this limitation and serves as a warning against interpreting our results as “OpenAlex is now just like Scopus”.

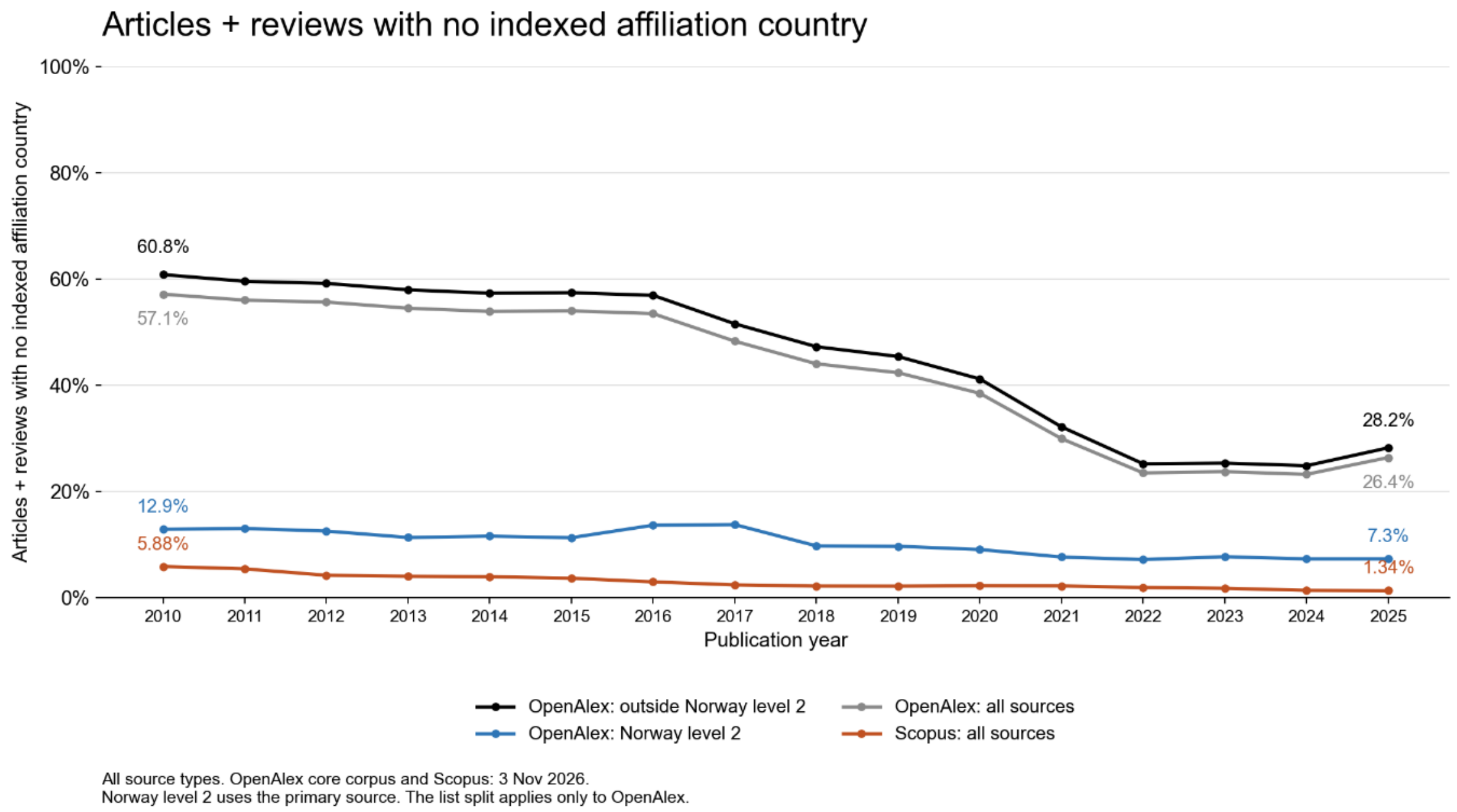

Figure 11. Shares of articles and reviews without country assignment in Scopus (all sources) and OpenAlex (all sources, Norway List second level sources, and sources outside it).

Continued efforts to improve the completeness and precision of institutional affiliations, including collaboration with national research systems and institutional users, and, of course, publishers will be essential to realizing OpenAlex’s full potential as an open alternative for research evaluation.

**Acknowledgments**

Data in this study were partly obtained from the German Competence Network for Bibliometrics (https://bibliometrie.info/), funded by the German Federal Ministry of Research, Technology and Space (Grant: 16WIK2501A).

**Open code and data**

Freely available and fully documented for replication available at https://github.com/IvanSterligov/Oalex_vs_Scp_orgs_journals Replication requires obtaining your own OpenAlex and Scopus API keys, available for free. Examples of manually checked VoRs (PDFs in OA) are included in the Online Supplement pack.

**AI usage disclosure**

We used OpenAI Codex (mainly with Sol on high and medium) and occasionally Claude (with Opus 5 and 5.5) to design the data collection, analysis and Matplotlib and ggplot2 figure generation scripts according to our specifications. Codex was also used for Online Supplement text initial generation, and for language editing and condensing in parts of the main manuscript. Authors are fully responsible for the resulting submission.

**Author contributions**

AL: Conceptualization, Formal analysis, Investigation, Methodology, Visualization, Writing – original draft, Writing – review & editing.

IS: Conceptualization, Data curation, Formal analysis, Investigation, Methodology, Software, Visualization, Writing – original draft, Writing – review & editing.

Appendix 1. Distribution of the institution-level relative differences and correlations between the institutions' OpenAlex and Scopus number of publications by publication year

| Year | Inst. | % difference, summed counts | Institutions' % difference, 100 × (OpenAlex − Scopus)/Scopus | | | Institutions by OpenAlex count | | | Correlation of the counts | |
|---|---|---|---|---|---|---|---|---|---|---|
| | | | Median | IQR | P10 to P90 | > Scopus + 10% | within ±10% | < Scopus − 10% | Spearman's rho | Lin's CCC |
| 2020 | 193 | −4.9 | -2.5 | -9.3 to +5.7 | -16.5 to +16.7 | 17% | 60% | 23% | 0.984 | 0.993 |
| 2021 | 198 | −5.5 | -3.4 | -9.9 to +4.9 | -17.9 to +18.8 | 13% | 64% | 23% | 0.990 | 0.992 |
| 2022 | 196 | −7.6 | -5.0 | -11.5 to +0.3 | -18.2 to +10.9 | 10% | 60% | 30% | 0.988 | 0.988 |
| 2023 | 195 | −7.8 | -4.0 | -11.8 to +2.7 | -17.9 to +10.5 | 10% | 59% | 31% | 0.990 | 0.988 |
| 2024 | 196 | −8.2 | -5.9 | -11.4 to +1.9 | -19.5 to +13.0 | 12% | 59% | 29% | 0.991 | 0.988 |
| 2025 | 196 | −6.6 | -3.8 | -12.0 to +5.1 | -21.2 to +14.5 | 14% | 56% | 31% | 0.990 | 0.989 |

Appendix 2. Distribution of the institution-level relative differences and correlations between the institutions' OpenAlex and Scopus number of publications by research field

| Field | Inst. | Jour. | % difference, summed counts | Institutions' % difference, 100 × (OpenAlex − Scopus)/Scopus | | | Institutions by OpenAlex count | | | Correlation of the counts | |
|---|---|---|---|---|---|---|---|---|---|---|---|
| | | | | Median | IQR | P10 to P90 | > Scopus + 10% | within ±10% | < Scopus − 10% | Spearman's rho | Lin's CCC |
| Natural Sciences and Engineering | | | | | | | | | | | |
| General Technology | 73 | 6 | −22.3 | -22.2 | -33.3 to -15.0 | -42.9 to +0.0 | 3% | 18% | 79% | 0.948 | 0.941 |
| Electric power engineering | 13 | 4 | −8.0 | -10.0 | -15.7 to -5.9 | -33.3 to +0.0 | 0% | 54% | 46% | 0.985 | 0.980 |
| Mathematics | 94 | 54 | −7.9 | -7.9 | -13.3 to +0.0 | -19.7 to +4.8 | 5% | 51% | 44% | 0.995 | 0.990 |
| Physics | 109 | 21 | −7.8 | -2.9 | -8.6 to +0.0 | -12.6 to +5.6 | 6% | 77% | 17% | 0.995 | 0.981 |
| Biology | 136 | 47 | −5.9 | -0.3 | -4.2 to +2.3 | -15.0 to +16.7 | 11% | 74% | 15% | 0.977 | 0.957 |
| Biotechnology | 38 | 4 | −4.4 | +0.0 | -8.3 to +0.0 | -16.7 to +6.7 | 5% | 71% | 24% | 0.978 | 0.990 |
| Chemical Engineering | 62 | 4 | −3.8 | +0.0 | -7.5 to +0.0 | -14.3 to +7.7 | 6% | 73% | 21% | 0.987 | 0.996 |
| Earth sciences | 100 | 27 | −2.9 | -1.7 | -4.0 to +0.0 | -11.1 to +6.6 | 4% | 85% | 11% | 0.985 | 0.997 |
| Interdisciplinary Natural Sciences | 168 | 6 | −2.5 | +0.0 | -5.0 to +3.0 | -15.0 to +10.5 | 10% | 76% | 14% | 0.981 | 0.990 |
| Computer Science and Informatics | 90 | 37 | −2.4 | +0.0 | -5.7 to +2.0 | -12.5 to +12.5 | 11% | 74% | 14% | 0.991 | 0.998 |
| Materials Science and Engineering | 101 | 16 | −2.3 | -1.8 | -5.7 to +0.0 | -9.5 to +11.1 | 11% | 79% | 10% | 0.996 | 0.996 |
| Electronics and cybernetics | 90 | 7 | −2.1 | +0.0 | -10.0 to +0.0 | -25.0 to +9.1 | 8% | 68% | 24% | 0.967 | 0.990 |
| Mechanical engineering | 45 | 11 | −2.1 | +0.0 | -6.2 to +1.8 | -14.3 to +12.5 | 16% | 73% | 11% | 0.990 | 0.997 |
| Civil Engineering | 31 | 8 | −1.7 | +0.0 | -9.1 to +0.0 | -12.5 to +4.1 | 6% | 71% | 23% | 0.981 | 0.999 |
| Energy | 49 | 6 | −1.7 | +0.0 | -4.9 to +0.0 | -16.7 to +12.5 | 10% | 73% | 16% | 0.988 | 0.999 |
| Chemistry | 113 | 13 | +0.1 | +0.0 | -1.7 to +3.1 | -5.3 to +16.7 | 13% | 82% | 4% | 0.996 | 0.997 |
| Health Sciences | | | | | | | | | | | |
| Haematology | 46 | 3 | −36.3 | -36.3 | -56.2 to -15.4 | -82.9 to -4.5 | 4% | 15% | 80% | 0.826 | 0.754 |
| Oncology | 49 | 7 | −20.0 | -19.5 | -32.4 to +0.0 | -56.0 to +20.0 | 10% | 29% | 61% | 0.901 | 0.930 |
| Cardiovascular and Respiratory Systems | 49 | 6 | −18.3 | -15.8 | -39.1 to -4.6 | -53.1 to +15.4 | 10% | 22% | 67% | 0.938 | 0.958 |
| Radiology, Nuclear Medicine and Medical Imaging | 41 | 4 | −17.9 | -13.7 | -34.1 to +0.0 | -61.5 to +5.9 | 10% | 32% | 59% | 0.900 | 0.917 |
| Infectious Diseases | 49 | 6 | −17.4 | -16.4 | -28.6 to -3.7 | -43.1 to +1.2 | 6% | 35% | 59% | 0.934 | 0.909 |
| General Medicine | 62 | 8 | −15.8 | -8.0 | -25.0 to +0.0 | -48.6 to +11.1 | 11% | 42% | 47% | 0.948 | 0.913 |
| Surgical Sciences | 35 | 8 | −12.5 | -21.4 | -37.5 to +0.0 | -50.0 to +5.6 | 3% | 37% | 60% | 0.903 | 0.964 |
| Neurology | 76 | 7 | −10.9 | -5.9 | -19.0 to +0.0 | -46.2 to +7.4 | 5% | 54% | 41% | 0.953 | 0.976 |
| Psychiatry | 64 | 8 | −9.1 | -6.2 | -15.4 to +0.0 | -31.2 to +14.3 | 14% | 55% | 31% | 0.976 | 0.983 |
| Sport Sciences | 12 | 3 | −8.8 | -5.0 | -14.3 to +7.7 | -25.0 to +12.5 | 17% | 42% | 42% | 0.770 | 0.927 |
| Pharmacology and Toxicology | 72 | 14 | −7.4 | -5.9 | -11.8 to +0.0 | -18.2 to +16.7 | 14% | 54% | 32% | 0.984 | 0.980 |
| Public, Environmental and Occupational Health | 90 | 22 | −6.4 | +0.0 | -11.3 to +0.0 | -25.7 to +9.4 | 9% | 64% | 27% | 0.974 | 0.980 |
| Psychology | 68 | 17 | −5.6 | -2.6 | -11.1 to +0.0 | -18.2 to +14.3 | 10% | 65% | 25% | 0.974 | 0.981 |
| Biomedicine | 130 | 39 | −4.3 | +0.0 | -8.8 to +0.0 | -18.2 to +9.2 | 8% | 71% | 21% | 0.977 | 0.992 |
| Endocrinology | 43 | 5 | −3.6 | -4.9 | -12.0 to +0.0 | -24.1 to +10.9 | 12% | 63% | 26% | 0.950 | 0.967 |
| Veterinary Medicine | 14 | 4 | +1.5 | +0.0 | +0.0 to +10.0 | -12.5 to +41.7 | 21% | 64% | 14% | 0.936 | 0.998 |
| Social Science | | | | | | | | | | | |
| Economics | 56 | 12 | −9.9 | -5.7 | -11.1 to +0.0 | -20.8 to +9.1 | 9% | 59% | 32% | 0.946 | 0.818 |
| Business and Finance | 69 | 30 | −7.1 | -5.1 | -10.5 to +0.0 | -19.0 to +4.8 | 6% | 68% | 26% | 0.985 | 0.981 |
| Geography | 50 | 7 | −5.0 | +0.0 | -12.5 to +0.0 | -25.0 to +11.1 | 14% | 60% | 26% | 0.957 | 0.976 |
| Political Science | 29 | 15 | −3.6 | +0.0 | -12.5 to +0.0 | -21.7 to +12.5 | 14% | 59% | 28% | 0.977 | 0.976 |
| Sociology | 32 | 8 | −3.6 | +0.0 | -14.3 to +3.3 | -27.3 to +11.8 | 16% | 56% | 28% | 0.970 | 0.988 |

| | | | | | | | | | | |
|---|---|---|---|---|---|---|---|---|---|---|
| Education and Educational Research | 39 | 17 | −0.6 | +0.0 | -11.1 to +6.2 | -14.3 to +20.0 | 15% | 59% | 26% | 0.938 | 0.975 |
| Interdisciplinary Social Sciences | 16 | 7 | −0.4 | +0.0 | +0.0 to +14.3 | -9.1 to +20.0 | 38% | 56% | 6% | 0.927 | 0.977 |
| Library and Information Science | 25 | 10 | +6.8 | +0.0 | +0.0 to +0.0 | -8.3 to +20.0 | 24% | 68% | 8% | 0.902 | 0.975 |
| Humanities | | | | | | | | | | | |
| Philosophy and History of Ideas | 18 | 7 | −2.4 | -6.6 | -9.5 to +10.0 | -18.2 to +25.0 | 22% | 61% | 17% | 0.931 | 0.993 |
| Linguistics | 24 | 8 | −0.5 | +0.0 | -4.0 to +0.0 | -16.7 to +10.0 | 8% | 75% | 17% | 0.964 | 0.989 |
| Media and Communication | 20 | 8 | +6.6 | +0.0 | -12.5 to +18.2 | -33.3 to +900.0 | 40% | 35% | 25% | 0.513 | 0.899 |

## Appendix 3. Distribution of the institution-level relative differences and correlations between the institutions' OpenAlex and Scopus number of publications by publisher

| Publisher | Inst. | Jour. | % difference, summed counts | Institutions' % difference, 100 × (OpenAlex − Scopus)/Scopus | | | Institutions by OpenAlex count | | | Correlation of the counts | |
|---|---|---|---|---|---|---|---|---|---|---|---|
| | | | | Median | IQR | P10 to P90 | > Scopus + 10% | within ±10% | < Scopus − 10% | Spearman's rho | Lin's CCC |
| Optica Publishing Group (formerly OSA) | 64 | 3 | −26.8 | -29.1 | -37.5 to -20.0 | -44.4 to -7.4 | 2% | 12% | 86% | 0.949 | 0.892 |
| Society for Industrial and Applied Mathematics | 56 | 8 | −26.0 | -25.0 | -33.3 to -18.8 | -40.9 to +0.0 | 2% | 14% | 84% | 0.958 | 0.861 |
| Wolters Kluwer Health | 48 | 12 | −17.5 | -9.5 | -33.3 to +0.0 | -46.2 to +4.8 | 6% | 48% | 46% | 0.906 | 0.910 |
| American Medical Association | 48 | 9 | −17.0 | -18.2 | -31.2 to -2.4 | -50.0 to +25.0 | 12% | 27% | 60% | 0.936 | 0.949 |
| American Astronomical Society | 51 | 3 | −11.1 | -7.1 | -14.3 to +0.0 | -22.2 to +0.6 | 8% | 55% | 37% | 0.965 | 0.988 |
| Elsevier | 189 | 129 | −9.2 | -4.3 | -12.0 to +0.0 | -21.2 to +10.0 | 10% | 61% | 29% | 0.989 | 0.985 |
| Institute of Physics Publishing | 102 | 7 | −8.7 | -5.9 | -12.0 to +0.0 | -18.7 to +5.1 | 9% | 63% | 28% | 0.976 | 0.976 |
| American Society for Microbiology | 60 | 4 | −8.6 | -8.9 | -15.4 to +0.0 | -25.5 to +0.0 | 7% | 52% | 42% | 0.967 | 0.968 |
| BMJ Publishing Group | 43 | 5 | −8.5 | -8.3 | -15.4 to +0.0 | -37.5 to +7.5 | 5% | 51% | 44% | 0.934 | 0.953 |
| John Wiley & Sons | 105 | 19 | −6.9 | -2.2 | -7.6 to +0.0 | -17.6 to +12.5 | 12% | 67% | 21% | 0.981 | 0.960 |
| IEEE | 67 | 13 | −6.3 | -5.0 | -12.9 to +0.0 | -27.5 to +7.1 | 9% | 61% | 30% | 0.974 | 0.996 |
| Springer Nature | 193 | 104 | −5.8 | +0.0 | -6.5 to +3.6 | -14.5 to +10.0 | 9% | 74% | 17% | 0.986 | 0.992 |
| Cambridge University Press | 60 | 10 | −4.1 | -3.8 | -9.5 to +0.0 | -16.7 to +5.9 | 8% | 70% | 22% | 0.984 | 0.995 |
| Oxford University Press | 114 | 31 | −3.6 | -1.9 | -10.0 to +3.8 | -20.0 to +13.3 | 15% | 61% | 25% | 0.980 | 0.990 |
| SAGE | 39 | 21 | −2.4 | +0.0 | -9.1 to +6.7 | -16.7 to +20.0 | 21% | 62% | 18% | 0.946 | 0.984 |
| Public Library of Science | 89 | 5 | −2.1 | +0.0 | -7.1 to +0.0 | -20.8 to +16.7 | 13% | 65% | 21% | 0.963 | 0.981 |
| Cold Spring Harbor Laboratory Press | 28 | 4 | −1.6 | +0.0 | -9.1 to +4.5 | -28.6 to +14.3 | 18% | 64% | 18% | 0.960 | 0.960 |
| Copernicus Publications | 65 | 4 | −1.5 | +0.0 | -2.1 to +1.3 | -14.3 to +6.9 | 5% | 85% | 11% | 0.995 | 0.994 |
| Institute of Mathematical Statistics | 17 | 5 | −1.3 | +0.0 | -5.9 to +11.1 | -14.3 to +20.0 | 29% | 53% | 18% | 0.967 | 0.973 |
| American Chemical Society | 131 | 19 | +0.2 | +0.0 | -2.5 to +2.9 | -4.9 to +11.9 | 11% | 86% | 2% | 0.995 | 0.997 |
| Taylor & Francis | 26 | 17 | +0.7 | +0.0 | -10.0 to +13.3 | -16.7 to +30.0 | 31% | 50% | 19% | 0.755 | 0.936 |
| American Physical Society | 98 | 5 | +2.1 | +0.0 | -1.2 to +2.7 | -10.0 to +14.1 | 11% | 80% | 9% | 0.992 | 0.962 |
| Association for Computing Machinery | 13 | 3 | +3.1 | +0.0 | +0.0 to +5.9 | -8.3 to +20.0 | 15% | 77% | 8% | 0.930 | 0.980 |